\documentclass[lettersize,journal]{IEEEtran}

\usepackage{cite}
\usepackage{easybmat}
\usepackage{stfloats}
\usepackage{amsmath,amssymb,amsfonts}
\usepackage{graphicx}
\usepackage{textcomp}
\usepackage{footmisc}
\usepackage{algorithm}
\usepackage{epstopdf}
\usepackage{balance}
\usepackage{caption}
\usepackage{subcaption}
\usepackage{algorithmic}
\usepackage[dvipsnames]{xcolor}
\usepackage[letterpaper, left=0.7in, right=0.7in, bottom=1in, top=0.7in]{geometry}
\usepackage{booktabs}
\DeclareFontFamily{U}{mathx}{}
\DeclareFontShape{U}{mathx}{m}{n}{ <-> mathx10 }{}
\DeclareSymbolFont{mathx}{U}{mathx}{m}{n}
\DeclareFontSubstitution{U}{mathx}{m}{n}

\DeclareMathAccent{\widecheck}{0}{mathx}{"71}

\newtheorem{thm}{Theorem}
\newtheorem{prop}{Proposition}

\newtheorem{defi}{Definition}

\newenvironment{proof}{ \paragraph*{Proof}}{\hfill$\square$}
\renewcommand{\baselinestretch}{.93} 
\newcommand{\mrm}[1]{\ensuremath{\text{#1}}}
\renewcommand{\vec}{\ensuremath{\mrm{vec}}}
\newcommand{\diag}{\ensuremath{\mrm{diag}}}
\newcommand{\rank}{\ensuremath{\mrm{rank}}}
\newcommand{\bs}[1]{\ensuremath{\boldsymbol{#1}}}
\newcommand{\comment}[1]{}

\renewcommand{\H}{\boldsymbol{H}}

\begin{document}

\title{Channel Orthogonalization with Reconfigurable Surfaces: General Models, Theoretical Limits, and Effective Configuration}
\author{Juan Vidal Alegr\'{i}a, Johan Thunberg, Ove Edfors\\
\thanks{This paper is built upon previous results presented at the 2022 IEEE GLOBECOM workshops \cite{globecom22}.\par This work was supported by the Swedish Foundation of Strategic Research (CHI19-0001) and by the Swedish Research Council (2019-04769).\par J. Vidal Alegr\'{i}a, J. Thunberg, and O. Edfors are with Lund University, SE-22100 Lund, Sweden (corresponding email: juan.vidal\_alegria@eit.lth.se).} }

\maketitle

\begin{abstract}
We envision a future in which multi-antenna technology effectively exploits the spatial domain as a set of non-interfering orthogonal resources, allowing for flexible resource allocation and efficient modulation/demodulation. We may refer to this paradigm as orthogonal space-division multiplexing (OSDM). On the other hand, reconfigurable intelligent surface (RIS) has emerged as a promising technology which allows shaping the propagation environment for improved performance. This paper studies the ability of three extended types of reconfigurable surface (RS), including the recently proposed beyond diagonal RIS (BD-RIS), to achieve perfectly orthogonal channels in a general multi-user multiple-input multiple-output (MU-MIMO) scenario. We consider practical implementations for the three types of RS consisting of passive components, and obtain the corresponding restrictions on their reconfigurability. We then use these restrictions to derive closed-form conditions and explicit expressions for achieving arbitrary (orthogonal) channels. We also study the problem of exploiting the degrees of freedom (DoFs) from the channel orthogonality constraint to maximize the channel gain while maintaining the passive RS constraints, and we propose some initial methods with satisfying performance. Finally, we provide some channel estimation and RS configuration techniques within this framework, where the computations are assumed to be performed at the BS, and we derive some limits on the amount of overhead required to achieve channel orthogonalization with RSs. The numerical results confirm the theoretical findings, showing that channel orthogonality with passive RSs can be effectively achieved in practical environments as long as the direct channel is not significant with respect to the RS cascaded channel. We thus take some important steps towards realizing OSDM.

\end{abstract}

\section{Introduction}\label{section:intro}
\IEEEPARstart{M}{odern} wireless communication systems tend to favor the use of modulations based upon orthogonal time-frequency resources, which may be employed independently without interfering each other. This can be seen in the popular orthogonal frequency division multiplexing (OFDM) \cite{ofdm}, considered in most current wireless communication standards, but also in more recent proposals such as orthogonal time frequency space (OTFS) \cite{otfs}, where the orthogonality is conversely achieved in the delay-Doppler domain. On the other hand, since the introduction of MIMO technology \cite{roy1996spatial}, the spatial domain has become a new resource which, in the same way as frequency or time, can be exploited for simultaneous transmission of independent streams of data. We may thus expect that efficient exploitation of the spatial domain may favor the use of techniques to divide it into a set of orthogonal resources, already hinted by the derivation of MIMO capacity \cite{telatar}. The extra challenge of the spatial domain comes from the fact that the propagation channel is scenario dependent, and we may have some limitations when it comes to channel estimation, as well as precoding/decoding \cite{mimo}.

\subsection{Motivation}
Multi-user MIMO (MU-MIMO) \cite{jindal} stands out as one of the main solutions for base station (BS) operation in modern wireless communication systems. Massive MIMO \cite{marzetta}, the large-scale version of MU-MIMO, has proved to be key in the development of 5G \cite{emil_next}, owing its fame to the improved spectral efficiency \cite{mMIMO_cap} and energy efficiency \cite{mMIMO_en_eff} from increasing the number of antennas at the BS. An important feature of massive MIMO is that, under rich multipath propagation, increasing the antennas at the BS leads to channel hardening and favorable propagation \cite{mmimo_book}. This means that the channel matrix becomes close to a semi-unitary matrix, allowing for enhanced multiplexing of user equipments (UEs) in the spatial domain at reduced complexity. Channel hardening has been observed to a fair extent in real measurements \cite{sara}; however, perfect channel orthogonality cannot be generally assured, specially when considering reduced array sizes or scarcely scattering environments. On the other hand, if we consider some of the technologies beyond massive MIMO under consideration for upcoming 6G \cite{emil_next,6g}, such as cell-free massive MIMO \cite{cf_mMIMO} or extra-large MIMO (XL-MIMO) \cite{xl-mimo}, channel hardening and favorable propagation may become even more compromised due to the non-stationarities of the channel when antennas have physically large separation \cite{ch_hard_cf_mMIMO,ch_hard_emil,xl-mimo}.  Moreover, the current interest in exploiting higher frequency bands, such as millimeter wave (mmWave) and terahertz (THz) \cite{6g}, reduces the validity of the rich scattering assumption due to the limited reflections and high insertion losses at these bands \cite{mmwave_meas,thz_meas}. {Hence, the channel orthogonality of mmWave channels is further reduced, limiting its spatial multiplexing capabilities \cite{mmWave_orth}.}

Reconfigurable intelligent surface (RIS) \cite{en_eff,RIS}, also known as intelligent reflective surface (IRS), is a technology with potential to become a key enabler for next generation communication systems \cite{6g,RIS}. RIS provides energy-efficient control over the propagation channel \cite{en_eff,RIS}, since it typically consists of a large number of passive reconfigurable elements whose reflection coefficients may be conveniently selected to generate adaptable reflections. Much of the research on RIS has focused on its impressive beamforming gains \cite{ris_scaling_law,RIS}. However, RIS may also be employed for improving the multiplexing capabilities in general MIMO scenarios. For example, \cite{rank_impr,rank2} proposes to employ the RIS reflection to increase the effective rank of a MIMO channel, where \cite{rank2} further provides experimental validation of this use case. Moreover, RIS optimizations based on capacity-related metrics also lead to improved multiplexing performance under certain scenarios \cite{sum_rate, rate}. 

{Channel orthogonality of the MU-MIMO channel is the most desirable property when it comes to performing spatial multiplexing \cite{mimo,mmWave_orth,osdm}. Thus, it feels of utmost interest to explore methods that allow converging towards orthogonal spatial domain multiplexing (OSDM), which may understood as the spatial counterpart of OFDM. Early work in \cite{osdm} considered the fundamental problem of orthogonalizing downlink MU-MIMO channels by jointly designing the BS precoder and the UEs equalizers. However, the considered framework had very limited practicality since it depended on multi-antenna UEs with full downlink channel state information (CIS) knowldege (including the CSI to other UEs). In the context of mmWave and THz technologies, the angular domain \cite{THz_orth}, as well as the delay-Doppler domain \cite{otfs}, offer interesting alternatives to isolate orthogonal spatial resources. However, the sparsity of the channel in these domains may limit substantially the number of streams that can be simultaneously multiplexed.}

In this paper, we address the fundamental task of enforcing channel orthogonality by employing RIS-related technology. Previous work in \cite{int_nulling}, proposed to use RIS to eliminate the interference between transceiver pairs, which leads to a specific form of channel orthogonality. However, \cite{int_nulling} focuses on an perfect CSI scenario with direct channel blockage, and where the cascaded channel through the RIS consists of a pure line-of-sight (LoS) channel, which may limit the validity of the results in more general scenarios. Note also that channel estimation in RIS is one of the main challenges faced by this technology due to the large pilot overhead required to estimate the cascaded channel coefficients \cite{ris_ch_est}. Furthermore, in MU-MIMO we may have some extra degrees of freedom (DoFs) by allowing for more general orthogonal MIMO channels \cite{globecom23} other than the identity channel enforced in \cite{int_nulling}.

Some extended reconfigurable surface (RS) technologies have been studied in the literature beyond RIS. For example, \cite{bd-ris,bd-ris1} consider the use of beyond diagonal RIS (BD-RIS), which consists of a generalized model of RIS where the reconfigurable elements may be interconnected through reconfigurable impedance networks, leading to an arbitrary reciprocal network. Due to its extra reconfigurability options, BD-RIS offers improved performance in terms of beamforming gain, up to 62\% \cite{bd-ris}. Thus, it is natural to expect that BD-RIS will also outperform RIS in the task of orthogonalizing the MIMO channel, which, to the best of our knowledge, has not been studied. In \cite{globecom22}, two further extended RS models are considered, amplitude-reconfigurable intelligent surface (ARIS), which corresponds to a variant of RIS which allows for amplitude reconfigurability (also contemplated to some extent in \cite{RIS} and \cite{bd-ris}), and fully-reconfigurable intelligent surface (FRIS), which considers the upper limit of having an unrestricted reflection matrix. In fact, considering the passive restriction, as well as the direct channel blockage assumption, the FRIS model leads to the capacity achieving reflection matrix derived in \cite{bartoli}, which is given by an unconstrained unitary matrix. In this work, we will consider some of these RS technologies in the task of MU-MIMO channel orthogonalization towards OSDM.

\subsection{Contribution}
This paper extends the results from \cite{globecom22}, where the problem of channel orthogonalization was preliminarily addressed considering ARIS and FRIS. However, in \cite{globecom22} we did not give a formal treatment of the passive constraints of ARIS and FRIS, since we considered constraining the average RS power instead of completely avoiding the use of active elements. In \cite{globecom23}, we employed a similar framework, but we instead focused on deriving the extra DoFs for data transmission available in the RS-BS link, which we now disregard. In this work, we study the problem of MU-MIMO channel orthogonalization based on purely passive RS technology. We include in our analysis the BD-RIS model, corresponding to the reciprocal version of FRIS, which was not considered in \cite{globecom22}. To the best of our knowledge, the use of BD-RIS, or generalized RS techonology based on purely passive components, has not been previously studied in the context of channel orthogonalization. The main contributions may be summarized as follows:
\begin{itemize}
    \item We provide formal restrictions on the reflection matrix of ARIS, FRIS, and BD-RIS, connecting them to possible implementations based on reconfigurable impedance networks.
    \item We derive necessary conditions on the number of elements, and corresponding reflection matrix, for selecting arbitrary (orthogonal) channels using ARIS, FRIS, and BD-RIS. These conditions are also sufficient when relaxing the passive constraint, or when the direct channel is blocked. In \cite{globecom22} we only included achievable bounds for ARIS and FRIS with relaxed passive constraint.
    \item We provide methods for orthogonal channel selection with ARIS, FRIS, and BD-RIS maximizing channel gain while fulfilling the passive constraint. Some methods employ similar tools as those in \cite{globecom22}, but the main channel selection algorithm is novel. Moreover, we now enforce a strict passive constraint unlike \cite{globecom22}, where a relaxed version based on average power was considered.
    \item We provide generalized methods for channel estimation and RS configuration with passive ARIS, FRIS, and BD-RIS, as well as theoretical bounds on the required channel estimation overhead. In \cite{globecom22}, a similar method was provided for ARIS, but for FRIS it was assumed that the RS could send pilots, which we hereby avoid.
\end{itemize}

\subsection{Paper Organization and Notation}
The rest of the paper is organized as follows. Section~\ref{section:model} presents the system model, including a discussion about implementation and restrictions of the different RS models. Section~\ref{sec:orth} provides formal theoretical results on the achievability of arbitrary (orthogonal) channels with RS. Section~\ref{sec:min} studies orthogonal channel selection, and provides an algorithm to perform this task. Section~\ref{sec:ch} presents a general channel estimation and RS configuration method for (orthogonal) channel selection. Section~\ref{section:num_res} includes numerical results to evaluate the presented theory. Finally, Section~\ref{section:num_res} concludes the paper with some final remarks and future work.

In this paper, lowercase, bold lowercase and bold uppercase letters stand for scalars, column vectors and matrices, respectively. The operations $(\cdot)^{\text{T}}$, $(\cdot)^*$, $(\cdot)^{\text{H}}$, and $(\cdot)^{\dagger}$ denote transpose, conjugate, conjugate transpose, and pseudoinverse, respectively. The trace operator is written as $\mrm{tr}(\cdot)$. The operation $\mathrm{diag}(\bs{a})$ gives a diagonal matrix with the input vector $\bs{a}$ as diagonal elements. $\bs{A} \otimes \bs{B}$ denotes the Kronecker product between matrices $\bs{A}$  and $\bs{B}$. $\mathbf{I}_i$ corresponds to the identity matrix of size $i$, $\boldsymbol{1}_{i\times j}$ denotes the $i \times j$ all-ones matrix,  and $\boldsymbol{0}_{i \times j}$ denotes the $i \times j$ all-zeros matrix (the subindices may be omitted if they can be derived from the context). $\{\boldsymbol{A}\}_{i,j}$ is the $(i,j)$th element of $\boldsymbol{A}$. $\boldsymbol{E}_{i,j}$ is a matrix having a single non-zero element in position $(i,j)$ equal to 1 (the dimensions are given by the context).

\section{System model}\label{section:model}
Let us consider a scenario where $K$ UEs are communicating in the uplink with an $M$-antenna BS through a narrowband channel. The communication link is assisted by an $N$-element RS. The received signal may be expressed in complex baseband as an $M$-sized vector given by
\begin{equation}\label{eq:ul_model}
\bs{y} = \bs{H}\bs{s} + \bs{n},
\end{equation}
where $\bs{H}$ is the $M\times K$ channel matrix, $\boldsymbol{s}$ is the $K \times 1$ vector of complex baseband symbols transmitted by the UEs, with $\mathbb{E}(\bs{s}\bs{s}^\mrm{H})=E_s \mathbf{I}_K$, and $\boldsymbol{n}$ is the random noise vector modeled by $\bs{n}\sim \mathcal{CN}(\boldsymbol{0}_{M\times 1}, N_0\mathbf{I}_M)$. 

{We consider the general assumption $M\geq K$, but some of our results are straightforwardly applicable for $M<K$. Specifically, Propositions~\ref{prop:ARIS} and \ref{prop:FRIS}, as well as Theorem~\ref{th:BD-RIS}, can still be used to get the conditions on how to enforce an arbitrary $M\times K$ channel matrix. However, this work focuses on the ability to serve the $K$ UEs under orthogonal spatial resources, while the number of such resources is fundamentally limited by the rank of the channel matrix, which is limited by $\min(M,K)$. Hence, having $M\geq K$ is a fundamental prerequisite to be able to serve $K$ users under orthogonal spatial resources.}

In a general RS scenario, we may assume that there exists a direct link between the UEs and the BS, as well as a reflected channel through the RS. Thus, we may express the channel matrix as
\begin{equation}\label{eq:channel}
\H = \H_0+\H_1 \bs{\Theta} \H_2,
\end{equation}
where $\H_0$ corresponds to the $M\times K$ channel matrix associated to the direct link between the BS and the UEs, $\H_1$ corresponds to the $M\times N$ channel matrix associated to the BS-to-RS link, $\H_2$ corresponds to the channel matrix associated to the RS-to-UEs link, and $\bs{\Theta}$ is the reflection matrix, which characterizes the narrowband response of the RS for a given configuration.

\subsection{RS Models}
\begin{figure*}[h]
  \begin{subfigure}[h]{0.29\linewidth}
    \centering
    \includegraphics[scale=0.93]{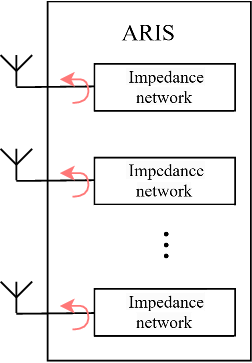}
\caption{}\label{fig:rs_arch_aris}
  \end{subfigure}
  \begin{subfigure}[h]{0.29\linewidth}
    \centering
    \includegraphics[scale=0.93]{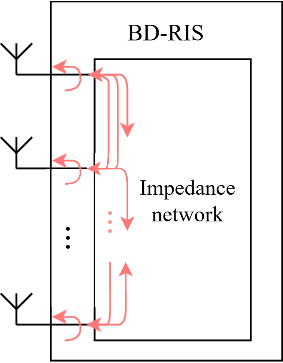}
\caption{}
\label{fig:rs_arch_bdris}
  \end{subfigure}
\begin{subfigure}[h]{0.29\linewidth}
    \centering
    \includegraphics[scale=0.93]{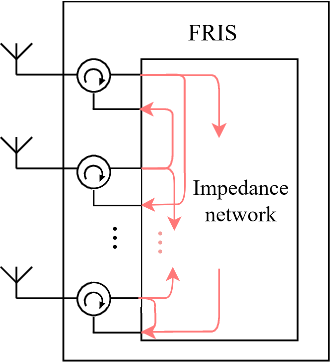}
\vspace{-1em}
\caption{}
\label{fig:rs_arch_fris}
  \end{subfigure}
\caption{RS models considered in this work.} 
\label{fig:rs_arch}
\vspace{-1em}
\end{figure*}

The most widespread model of RS is RIS, which considers that each RS element is associated to a passive network with controllable reflection coefficient. Furthermore, it is commonly assumed that this reflection coefficient corresponds to a pure phase shift, so that the power being reflected at the RIS is maximized and no energy is burnt at the passive network associated to each RIS element. This leads to the usual restriction on the narrowband reflection matrix
\begin{equation}\label{eq:mod_RIS}
    \bs{\Theta}_\mrm{RIS} = \diag\left(\exp(j\phi_1),\dots,\exp(j\phi_N)\right).
\end{equation}

The restriction put on RIS to reflect all the incoming power can become a limiting factor in certain situations where it may be interesting to sacrifice part of the reflected power with the aim of improving other qualities of the  channel, e.g., its multiplexing performance. Thus, we consider an RS model which generalizes that of RIS by relaxing the previous restriction and allowing amplitude reconfigurability at each RS element, together with the common phase reconfigurability.
\begin{defi}
    \textit{Amplitude reconfigurable intelligent surface (ARIS)} is hereby defined as an RS model whose reflection matrix satisfies
     \begin{equation}\label{eq:ARIS_cons}
        \bs{\Theta}_\mrm{ARIS} = \diag\left(\alpha_1,\dots,\alpha_N\right),\;\; \vert \alpha_i\vert^2\leq 1 \;\; \forall i.
    \end{equation}
    An \textit{ARIS with relaxed power constraint} corresponds to an ARIS where the restriction $\vert \alpha_i\vert^2\leq 1$ is disregarded.
\end{defi}

From \eqref{eq:ARIS_cons}, we may note that the RIS restriction, which can be written as $\vert\alpha_i \vert=1 \;\forall i$, is now relaxed to $\vert \alpha_i\vert^2\leq 1 \;\forall i$. This comes from considering fully passive impedance networks which may include reconfigurable resistive components, e.g., varistors, apart from the purely reactive ones commonly assumed in RIS literature. The unrestricted case can also be implemented by including active elements, e.g., amplifiers, leading to the ARIS with relaxed power constraint, which is also found in the literature under the term \textit{active RIS} \cite{ARIS_basar,ARIS_larsson}. However, it is important to point out that including amplification in an RS compromises the assumption of IID noise with fixed power in \eqref{eq:ul_model}, and correspondingly the optimality of enforcing channel orthogonalization for improved multiplexing. We thus limit our study to passive RS technologies, for which the system would exhibit optimal multiplexing performance under channel orthogonality \cite{mimo}. A potential implementation of ARIS is illustrated in Fig.~\ref{fig:rs_arch_aris}.

% Note that we have disregarded the restriction of having each $\alpha_i$ of amplitude $1$ associated to passive reflective networks. This may lead in some cases to requiring active amplification on top of the passive network, which has already been considered in the RIS context and some hardware implications are given in \cite{ARIS_basar,ARIS_larsson}. However, we will also propose methods to minimize these weights with the objective of removing the need of active amplification.
%Note that we may further ignore the restriction $\vert\alpha_i\vert^2\leq 1$ if we consider the use of active amplification on top of the passive network. This has already been considered in the RIS context, where some of the hardware implications to realize these systems are given in \cite{ARIS_basar,ARIS_larsson}. In fact, some our general results may consider this relaxation to enable mathematical tractability, but this may be dealt with by minimizing the resulting amplitudes as we will see.

Another RS model that has been studied in the literature is BD-RIS \cite{bd-ris,bd-ris1}, which extends the idea of RIS by allowing to interconnect different RS elements through a reconfigurable passive (impedance) network. The correspondence between the impendance network interconnecting the RS elements, and the resulting reflection matrix is one-to-one, and obtained from microwave network theory \cite{pozar,bd-ris} as
\begin{equation}\label{eq:ref_imp}
    \bs{\Theta} = (\bs{Z}+\mathbf{I}_N)^{-1}(\bs{Z}-\mathbf{I}_N),
\end{equation}
where $\bs{Z}$ is the $N \times N$ impedance matrix. Hence, the restrictions on the reflection matrix when considering different BD-RIS architectures have direct correspondence with the restrictions on the impendance network leading to $\bs{Z}$ \cite{bd-ris}. In the case of ARIS, which can be seen as a restricted BD-RIS, no interconnection is assumed between elements, leading to an $N$-port network with diagonal impedance matrix, which correspondingly gives a diagonal reflection matrix. The power constraint $\vert \alpha_i \vert^2\leq 1$ further comes from the fact that, in passive (impedance) networks, the resistive part can only be positive. { In the general case, a passive network should have an output power lower than or equal to the input power independent of the input signal. Thus, we should have  $\bs{x}^\mrm{H}\bs{\Theta}^\mrm{H}\bs{\Theta}\bs{x}\leq \Vert\bs{x}\Vert^2$, which, by the Rayleigh quotient \cite{mat_anal}, leads to the restriction on the largest eigenvalue of $\bs{\Theta}^\mrm{H}\bs{\Theta}$ (i.e., the spectral norm of $\bs{\Theta}$) being smaller than or equal to $1$. In \cite{bd-ris}, the previous constraint is equivalently stated in the form $\bs{\Theta}^\mrm{H}\bs{\Theta}\preceq \mathbf{I}_N$.} Moreover, assuming reciprocal networks, e.g., pure impedance networks, the impedance matrix has the extra restriction that it is symmetric, which further leads to a symmetric constraint in the reflection matrix, i.e., $\bs{\Theta}=\bs{\Theta}^\mrm{T}$.

As happened with the RIS case, most of the literature on BD-RIS \cite{bd-ris,bd-ris1} focuses on the lossless case, i.e., where all the incident power at the BD-RIS is being reflected, leading to a unitary constraint on $\bs{\Theta}$. However, it may still be interesting, under certain scenarios, to sacrifice some reflected power at the BD-RIS for improving other channel properties. Hence, we will consider a generalization over the common BD-RIS \cite{bd-ris,bd-ris1} reflection model by assuming that the impedance network may include resistive (lossy) components instead of purely reactive (lossless) components. This leads to the following definition.
\begin{defi}
    \textit{Beyond diagonal reconfigurable surface (BD-RIS)} is hereby defined as an RS model whose reflection matrix satisfies
     \begin{equation}\label{eq:BD-RIS_cons}
        \bs{\Theta}_{\mrm{BD-RIS}} \in \{\bs{\Theta}\in \mathbb{C}^{N\times N}\;\vert \; \bs{\Theta}=\bs{\Theta}^{\mrm{T}},\, \Vert \bs{\Theta}\Vert^2_2 \leq 1\},
    \end{equation}
    where $\Vert \cdot \Vert_2$ corresponds to the matrix spectral norm. A \textit{BD-RIS with relaxed power constraint} corresponds to a BD-RIS where the restriction $\Vert \bs{\Theta}\Vert^2_2 \leq 1$ is disregarded.\footnote{\label{fn:amp}The restriction $\Vert \bs{\Theta}\Vert^2_2 \leq 1$ comes from the consideration of the passive (energy-efficient) nature of RS systems. Analogously as with ARIS, we may disregard such assumption by including amplification \cite{ARIS_larsson}.}
\end{defi}

Assuming a fully reconfigurable impedance network, an arbitrary reflection matrix fulfilling \eqref{eq:BD-RIS_cons} may be achieved by adjusting the respective impedance matrix in \eqref{eq:ref_imp}. {Note that ARIS also fulfills \eqref{eq:BD-RIS_cons} since the eigenvalues of $\bs{\Theta}_\mrm{ARIS}^\mrm{H}\bs{\Theta}_\mrm{ARIS}$ are directly given by $\vert\alpha_i\vert^2$, while a diagonal matrix is also symmetric. Moreover, we could equivalently express the ARIS passive constraint in terms of the spectral norm constraint in \eqref{eq:BD-RIS_cons}.} A potential implementation of BD-RIS is illustrated in Fig.~\ref{fig:rs_arch_bdris}. A more thorough discussion on specific implementations of the considered impedance networks is given in \cite{bd-ris1}.

We also consider an RS model corresponding to a further generalization of the previous models.
\begin{defi}
    \textit{Fully-reconfigurable intelligent surface (FRIS)} is hereby defined as an RS model whose reflection matrix satisfies
     \begin{equation}\label{eq:FRIS_cons}
    \bs{\Theta}_\mrm{FRIS}\in \{\bs{\Theta}\in \mathbb{C}^{N\times N}\;\vert \; \Vert \bs{\Theta}\Vert^2_2 \leq 1\}.
\end{equation}
    A \textit{FRIS with relaxed power constraint} corresponds to a FRIS where the restriction $\Vert \bs{\Theta}\Vert^2_2 \leq 1$ is disregarded.\footref{fn:amp}
\end{defi}

The practicality of FRIS is doubtful, but it naturally arises as a limit case for upper bounding the capabilities of RS systems. In fact, this type of RS gives the capacity achieving reflection matrix in \cite{bartoli}, where the non-symmetric reflection matrix is further restricted to be unitary, corresponding again to the lossless case, i.e., reflecting maximum power. In principle, a general reflection matrix, as given by \eqref{eq:FRIS_cons}, may be achieved by considering a passive non-reciprocal network, i.e., including elements such as passive circulators or isolators \cite{non-recip}. In Fig.~\ref{fig:rs_arch_fris} we illustrate a possible implementation based on circulators, where the impedance network has now $2N$ ports instead of $N$ as in the previous designs. Other possible implementations may use $N$ pairs of RS elements, half of them for receiving and the other half for transmitting, as well as isolators between both sides; however, such implementations may lead to a loss of channel reciprocity. In any case, the reader may take FRIS as a mere theoretical concept to understand the limits of RS systems.

\subsection{Note on the downlink scenario}
Assuming channel reciprocity, the results presented in this work may be trivially extended to the downlink scenario. Note that the architectures illustrated in Fig.~\ref{fig:rs_arch} would maintain the reciprocity of the channel for fixed RS configuration. Thus, the downlink channel would correspond to the transpose of the uplink channel, retaining the same enhanced qualities without the need for RS reconfiguration. In the case of channel orthogonality, the BS could simply use an MRT precoder to optimally multiplex the information sent to the different UEs, i.e., without incurring interference or noise enhancement.

\section{RS Channel Orthogonalization}\label{sec:orth}
%Modern wireless systems typically rely on modulations that allow us to divide the time-frequency domain into a set of orthogonal resources which may be used independently without having to worry about interfering other time-frequency resources---e.g., OFDM. This is of great convenience since it gives a way to multiplex in time and/or frequency independent streams of data. Such orthogonality property may be also desirable for the spatial domain, which is generally dependent on the propagation conditions. On the other hand, RS technology offers an effective way to improve the propagation channel according some metric, e.g., array gain, channel capacity, multiplexing gain, etc. Thus, it feels natural to employ RS technology to orthogonalize the spatial domain, so that an interference-free grid of space-time-frequency resources may be conveniently accessible.

The main goal of this work is to study how the previously defined RS technology may be used to enforce channel orthogonality in the spatial domain. This would allow us to perfectly multiplex the information intended to the different UEs (in both the uplink and the downlink), and treat each of them as parallel, non-interfering, and independent streams.\footnote{Note that, due to the passive nature of RS, we may assume that the covariance of the noise vector is unaffected by the reconfigurable reflection.} Moreover, this is achieved with the simplest precoder and equalization schemes, namely MRC and MRT, which are also extremely favorable in decentralized BS implementations \cite{latency}. We next define what is meant by channel orthogonality in the spatial domain.

\begin{defi}
We hereby define an \textit{orthogonal channel} as a wireless propagation channel leading to a channel matrix $\widetilde{\H}$ given by
\begin{equation}\label{eq:orth_ch}
\widetilde{\H}=\sqrt{\beta}\widetilde{\bs{U}}, 
\end{equation}
where $\beta$ is a real positive scalar corresponding to the \textit{channel gain}, and $\widetilde{\bs{U}}$ is a semi-unitary matrix, i.e., such that $\widetilde{\bs{U}}^\mrm{H}\widetilde{\bs{U}}=\mathbf{I}_K$. Alternatively we can say that $\widetilde{\bs{U}}$ is an element of the Stiefel manifold $\mathcal{S}(M,K)$, which may be thus constructed by taking $K$ columns of an $M\times M$ unitary matrix $\bs{U}\in \mathcal{U}(M)$.
\end{defi}

Note that the previous definition leads to a channel matrix whose squared singular values are all equal to $\beta$. We could consider a less restrictive definition by allowing for different eigenvalues while maintaining the orthogonality constraint, e.g., multiplying from the right in \eqref{eq:orth_ch} a $K\times K$ diagonal matrix. In fact, the results presented in this section are also directly applicable to that case. Nevertheless, we focus our exposition on orthogonal channels given by \eqref{eq:orth_ch} for notation simplicity and due to some increased benefits explained next.

Orthogonal channels, as given by \eqref{eq:orth_ch}, are hugely desirable in MU-MIMO systems for several reasons \cite{mimo}:
\begin{itemize}
    \item Full multiplexing gain is available since all of the singular values of the channel matrix are non-zero.
    \item The waterfilling algorithm \cite{telatar} is not needed for achieving capacity since all eigenvalues of the channel are equal.
    \item The sum-rate is equally distributed among the UEs since the orthogonal spatial streams have equal power.
    \item Simple linear equalization or precoding, namely MRC or MRT, achieves optimum performance, since it can exploit the orthogonal paths of the channel without the need for UE cooperation.
\end{itemize}

We will now study the requirements for the different RS models to achieve arbitrary channel matrices, and specifically, to achieve an arbitrary orthogonal channel. We will start by ignoring the RS power constraints for analytical tractability. Nevertheless, whenever the collection of channels $\bs{H}_0$, $\bs{H}_1$, and $\bs{H}_2$, as well as the desired channel $\widetilde{\bs{H}}$, allow for an RS configuration fulfilling the respective power constraint, the presented results will provide it. On the other hand, if said collection of channels leads to an RS configuration not fulfilling the power constraint, we will show in the next section how to employ the freedom in the orthogonality constraint to minimize the RS power such that it may be implemented using purely passive components. In the case of having $\bs{H}_0\approx \bs{0}_{M\times K}$, e.g., when the direct channel suffers from severe blockage, we will see that the RS power constrains may be always fulfilled by adjusting the channel gain.

%We next study how to define $\bs{\Theta}$, as well as what number of reconfigurable elements $N$ is required, in order to achieve arbitrary channels---and specifically orthogonal channels as defined in \eqref{eq:orth_ch}---considering the restrictions of ARIS, BD-RIS, and FRIS. Our initial results consider a relaxation over the constraints for the different RS models associated to having purely-passive (impedance) networks. One way to get around said constraint is to include some aplification in our design. However, another alternative which we will explore is to perform power minimization maintaining the channel orthogonality condition \eqref{eq:orth_ch} with the aim to reduce, and even remove altogether, the amplification requirement.

\subsection{FRIS}
We start by considering the FRIS model since this should lead to the most fundamental limits on the ability to generate arbitrary (orthogonal) channels with RS technology. The following proposition provides the conditions for FRIS to be able to generate arbitrary (orthogonal) channels.

\begin{prop}\label{prop:FRIS}
Given an arbitrary direct channel $\bs{H}_0$, and arbitrary full-rank channel matrices $\bs{H}_1$ and $\bs{H}_2$, a FRIS with relaxed power constraint is able to generate an arbitrary channel $\widetilde{\bs{H}}\neq \bs{H}_0$, and specifically an arbitrary orthogonal channel given by \eqref{eq:orth_ch}, if and only if
\begin{equation}\label{eq:cond_FRIS}
    N \geq \max(M,K).
\end{equation}
Said channel $\widetilde{\bs{H}}$ is achieved by configuring the FRIS reflection matrix as
\begin{equation}\label{eq:FRIS_conf}
    \bs{\Theta}_\mrm{FRIS} = \vec^{-1} \big(\bs{\mathcal{H}}^\dagger \bs{c}\big),
\end{equation}
where $\bs{\mathcal{H}}^\dagger$ corresponds to the right pseudoinverse of $\bs{\mathcal{H}}=(\bs{H}_2^\mrm{T}\otimes \bs{H}_1)$, and $\bs{c}=\vec(\widetilde{\bs{H}}-\bs{H}_0)$. An alternative (more compact) expression may be given by
\begin{equation}\label{eq:FRIS_conf_alt}
    \bs{\Theta}_\mrm{FRIS} = \bs{H}_1^\dagger\big(\widetilde{\bs{H}}-\bs{H}_0\big)\bs{H}_2^\dagger,
\end{equation}
where $\bs{H}_1^\dagger$ and $\bs{H}_2^\dagger$ correspond to the right and left pseudoinverses of $\bs{H}_1$ and $\bs{H}_2$, respectively.
\begin{proof}
We want to study solutions to the matrix equation
\begin{equation}\label{eq:FRISeq}
    \bs{H}_0+\bs{H}_1\bs{\Theta}\bs{H}_2 = \widetilde{\bs{H}},
\end{equation}
which corresponds to a linear system of equations where $\bs{\Theta}$ is the unrestricted $N\times N$ complex matrix of unknowns. If we move $\bs{H}_0$ to the right-hand side (RHS) and vectorize, we may express \eqref{eq:FRISeq} as
\begin{equation}\label{eq:FRIS_lineq}
    \bs{\mathcal{H}} \,\vec (\bs{\Theta}) = \bs{c},
\end{equation}
where $\bs{c}=\vec (\widetilde{\bs{H}}-\bs{H}_0)$ is a non-zero $MK$-sized vector (since $\widetilde{\bs{H}}\neq\bs{H}_0$), and $\bs{\mathcal{H}}= (\bs{H}_2^\mrm{T} \otimes \bs{H}_1)$ is an $MK \times N^2$ matrix. From the properties of the Kronecker product \cite{kron_prod}, we can characterize the rank of $\bs{\mathcal{H}}$ as
\begin{equation}
    \rank (\bs{\mathcal{H}})=\rank(\bs{H}_2)\cdot \rank(\bs{H}_1),
\end{equation}
where, since $\bs{H}_1$ and $\bs{H}_2$ are full rank, we have that $\rank(\bs{H}_1)=\min(M,N)$ and $\rank(\bs{H}_2)=\min(K,N)$. On the other hand, by fundamental linear algebra arguments, a solution to \eqref{eq:FRIS_lineq} can be found if and only if $\rank(\bs{\mathcal{H}})=M\cdot K$, which leads to the condition \eqref{eq:cond_FRIS}. Moreover, if a solution to \eqref{eq:FRIS_lineq} exists, it is given by
\begin{equation}
    \vec(\bs{\Theta})=\bs{\mathcal{H}}^\dagger \bs{c},
\end{equation}
where $\bs{\mathcal{H}}^\dagger$ is the right pseudoinverse of $\bs{\mathcal{H}}$. Note that there may be multiple solutions since the pseudoinverse may include $(N^2-MK)$ columns arbitrarily selected from the null-space of $\bs{\mathcal{H}}$. After inverse vectorization we reach \eqref{eq:FRIS_conf}. On the other hand, exploiting the property of the Kronecker product $(\bs{A}\otimes \bs{B})^\dagger=\bs{A}^\dagger\otimes \bs{B}^\dagger$ \cite{kron_prod} leads to \eqref{eq:FRIS_conf_alt}.
\end{proof}
\end{prop}

Proposition~\ref{prop:FRIS} sets a requirement on $N$ to generate arbitrary channels using FRIS. However, as seen in the previous section, practical implementations of FRIS may actually employ an impedance network with $2N$ ports. We will find out that, in some cases, this consideration could make FRIS implementations more restrictive than BD-RIS implementations.

\subsection{BD-RIS}
The following theorem delimits the capabilities of BD-RIS to achieve arbitrary (orthogonal) channels.

\begin{thm}\label{th:BD-RIS}
Given an arbitrary direct channel $\bs{H}_0$, and \textit{randomly chosen}\footnote{A \textit{randomly chosen} matrix is hereby defined as a matrix whose elements may have been drawn from arbitrary continuous distributions such that any submatrix of it is full-rank with probability 1, e.g., a realization of a Gaussian matrix with full-rank correlation.} channel matrices $\bs{H}_1$ and $\bs{H}_2$, a BD-RIS with relaxed power constraint is able to generate an arbitrary channel $\widetilde{\bs{H}}\neq \bs{H}_0$, and specifically an arbitrary orthogonal channel as given by \eqref{eq:orth_ch}, if and only if
\begin{equation}\label{eq:cond_BD-RIS}
    N \geq M+K-1.
\end{equation}
Furthermore, \eqref{eq:cond_BD-RIS} is also a sufficient condition for arbitrary full-rank $\bs{H}_1$ and $\bs{H}_2$. Said channel $\widetilde{\bs{H}}$ is achieved by configuring the BD-RIS reflection matrix as
\begin{equation}\label{eq:BD-RIS_conf}
\bs{\Theta}_\mrm{BD-RIS}=\vec^{-1}\Big(\big(\bs{K}^{(N,N)}+\mathbf{I}_{N^2}\big)\bs{Z}_\mrm{U/L}\big(\bs{\mathcal{H}}_\mrm{U}+\bs{\mathcal{H}}_\mrm{L}\big)^\dagger \bs{c}\Big)
\end{equation}
where $\bs{K}^{(N,N)}$ is the commutation matrix for $N\times N$ matrices \cite{comm_mtx}, $\bs{c}=\vec(\widetilde{\bs{H}}-\bs{H}_0)$, $\bs{\mathcal{H}}_\mrm{U/L}$ corresponds to an $N^2\times N(N+1)$ matrix constructed from the columns of $\bs{\mathcal{H}}$ in \eqref{eq:FRIS_conf} associated to the upper/lower triangular elements after vectorization, and $\bs{Z}_\mrm{U/L}$ is an $N^2\times N(N+1)/2$ matrix that pads zeros in the entries associated to elements below/above the diagonal after inverse vectorization. 
\begin{proof}
    See Appendix.
\end{proof}
\end{thm}
We may now note that, if we consider a FRIS implementation requiring a $2 N$-port impendance network (as illustrated in Fig.~\ref{fig:rs_arch_fris}), the minimum $N$ from \eqref{eq:cond_BD-RIS} and \eqref{eq:cond_FRIS} would translate into a stricter requirement in terms of impedance network ports for FRIS than for BD-RIS, giving an increase of $\max\big(M-K+1,K-M+1\big)$ ports. This may be understood by the fact that such FRIS implementation corresponds to a non-reciprocal version of a BD-RIS with $2 N$ elements where the non-reciprocicity actually reduces the DoFs. However, the reduction in antenna elements may still be desirable in some scenarios. On the other hand, from \eqref{eq:cond_BD-RIS} we may also notice that, assuming $M>1$ and/or $K>1$, BD-RIS does not make perfect use of all the $N(N+1)/2$ free variables in $\bs{\Theta}_\mrm{BD-RIS}$, since it actually requires an excess of $\lceil(M^2-M+K^2-K)/2\rceil$ free variables to be able to solve the $MK$ unknowns associated to the channel update.

\subsection{ARIS}
The following proposition particularizes the previous results to the ARIS restriction.

\begin{prop}\label{prop:ARIS}
Given an arbitrary direct channel $\bs{H}_0$, and \textit{randomly chosen} channel matrices $\bs{H}_1$ and $\bs{H}_2$, an ARIS with relaxed power constraint is able to generate an arbitrary channel $\widetilde{\bs{H}}\neq \bs{H}_0$, and specifically an arbitrary orthogonal channel given by \eqref{eq:orth_ch}, if and only if
\begin{equation}\label{eq:cond_ARIS}
    N \geq MK.
\end{equation}
Said channel $\widetilde{\bs{H}}$ is achieved by selecting the ARIS reflection matrix as
\begin{equation}\label{eq:ARIS_conf}
    \bs{\Theta}_\mrm{ARIS} = \diag \big(\bs{\mathcal{H}}_{D}^\dagger\bs{c}\big),
\end{equation}
where $\bs{c}=\vec(\widetilde{\bs{H}}-\bs{H}_0)$, and $\bs{\mathcal{H}}_{D}$ is an $MK \times N$ matrix constructed from the columns of $\bs{\mathcal{H}}$ from \eqref{eq:FRIS_conf} associated to the diagonal elements after vectorization.
\begin{proof}
We study the solutions to the equation
\begin{equation}\label{eq:RISA_eq}
    \H_0+\H_1 \bs{\Theta}_\mrm{D} \H_2 = \widetilde{\bs{H}},
\end{equation}
where $\bs{\Theta}_\mrm{D}=\diag (\bs{\alpha}_\mrm{D})$ for an $N$-sized vector of unknowns $\bs{\alpha}_\mrm{D}$. We may proceed by vectorizing as in the proof of Proposition~\ref{prop:FRIS}, leading to
\begin{equation}\label{eq:ARIS_lineq}
    \bs{\mathcal{H}}_\mrm{D} \bs{\alpha}_\mrm{D} = \bs{c},
\end{equation}
where $\bs{\mathcal{H}}_\mrm{D}$ is an $MK\times N$ matrix whose columns correspond to the columns of $\bs{\mathcal{H}}$ multiplying the diagonal elements of $\bs{\Theta}$ in \eqref{eq:FRIS_lineq}. Since $\bs{c}=\vec(\widetilde{\H}-\H_0)$ is a non-zero vector, \eqref{eq:ARIS_lineq} corresponds to a linear equation which is solvable if and only if $\rank(\bs{\mathcal{H}}_\mrm{D})=MK$. Let us assume \eqref{eq:cond_FRIS} since this is clearly a necessary condition for \eqref{eq:RISA_eq}. For arbitrary full rank matrices $\H_2$ and $\H_1$, we have that $\rank(\bs{\mathcal{H}})=MK$. On the other hand, if $\H_2$ and $\H_1$ are further \textit{randomly chosen}, any selection of columns/rows from $\bs{\mathcal{H}}$ will also be full rank with probability 1. Hence, having a selection $\bs{\mathcal{H}}_\mrm{D}$ of rank $MK$ is equivalent to \eqref{eq:cond_ARIS}. The solution \eqref{eq:ARIS_conf} is then trivially given by inverting $\bs{\mathcal{H}}_\mrm{D}$ in \eqref{eq:ARIS_lineq}, and constructing the diagonal matrix.

\end{proof}
\end{prop}

\section{Orthogonal channel selection}\label{sec:min}

In this section, we provide techniques for selecting suitable orthogonal channels such that the required reflection matrix fulfills the passive restrictions from \eqref{eq:FRIS_cons}, \eqref{eq:BD-RIS_cons}, and \eqref{eq:ARIS_cons}. In the previous section, we derived closed-form expressions for the reflection matrices required to obtain an arbitrary (orthogonal) channel with FRIS, ARIS, and BD-RIS, given in \eqref{eq:FRIS_conf}, \eqref{eq:BD-RIS_conf}, and \eqref{eq:ARIS_conf}, respectively. The idea now is to exploit the freedom in the orthogonality constraint to restrict the power of the reflection matrix such that no amplification is required at the respective RS models.

In the three RS models considered, the power constraint allowing for channel orthogonalization using only passive components may be expressed in terms of the spectral norm as
\begin{equation}\label{eq:gen_norm_cons}
    \Vert \bs{\Theta}_r\Vert^2_2  \triangleq \max_{\bs{x}} \frac{\bs{x}^\mrm{H}\bs{\Theta}_r^\mrm{H}\bs{\Theta}_r\bs{x}}{\Vert\bs{x}\Vert^2_2}\leq 1,
\end{equation}
where $r\in \{\mrm{FRIS}, \mrm{BD-RIS}, \mrm{ARIS}\}$, and $\bs{\Theta}_r$ may be obtained by substituting \eqref{eq:orth_ch}, in \eqref{eq:FRIS_conf}, \eqref{eq:BD-RIS_conf}, and \eqref{eq:ARIS_conf}. We may write $\bs{\Theta}_r$ as
\begin{equation}\label{eq:theta_gen}
    \bs{\Theta}_r = \vec^{-1}\Big(\widetilde{\bs{\mathcal{H}}}_r\big(\sqrt{\beta}\vec(\widetilde{\bs{U}})-\vec(\bs{H}_0)\big)\Big),
\end{equation}
where $\widetilde{\bs{\mathcal{H}}}_r$ is an $N^2\times MK$ full-rank matrix respectively given by
\begin{subequations}\label{eq:HH_r}
\begin{equation}
    \widetilde{\bs{\mathcal{H}}}_\mrm{FRIS} = \bs{\mathcal{H}}^\dagger
\end{equation}
\begin{equation}
    \widetilde{\bs{\mathcal{H}}}_\mrm{BD-RIS} = \big(\bs{K}^{(N,N)}+\mathbf{I}_{N^2}\big)\bs{Z}_\mrm{U/L}\big(\bs{\mathcal{H}}_\mrm{U}+\bs{\mathcal{H}}_\mrm{L}\big)^\dagger
\end{equation}
\begin{equation}
    \widetilde{\bs{\mathcal{H}}}_\mrm{ARIS} = \bs{Z}_\mrm{D}\bs{\mathcal{H}}_{D}^\dagger,
\end{equation} 
\end{subequations}
with $\bs{Z}_\mrm{D}$ corresponding to an $N^2\times N$ matrix that pads zeros in the entries associated to the off-diagonal elements after vectorization (analogue to $\bs{Z}_\mrm{U/L}$ for the upper/lower triangular elements). Note that, whenever $\bs{H}_0=\bs{0}$, the only term left in \eqref{eq:theta_gen} scales with $\sqrt{\beta}$, so if the direct channel is blocked we can always find a value of $\beta$ such that \eqref{eq:gen_norm_cons} is fulfilled. However, we are also interested in having the channel gain $\beta$ as high as possible since it multiplies the post-processed signal-to-noise ratio (SNR) for each UE, i.e., leading to a capacity per UE of $\log(1+\beta \mrm{SNR})$  \cite{inf_th}. In general, if a given $\widetilde{\bs{U}}$ allows for some $\beta \geq 0$ such that \eqref{eq:gen_norm_cons} is fulfilled, then $\Vert \bs{\Theta}_r\Vert^2_2$ is an increasing function in $\beta\geq \beta_0$ for some $\beta_0\geq 0$,\footnote{This can be seen by noting that $\Vert \bs{\Theta}_r\Vert^2_2$ is upper bounded by a quadratic expression in $\sqrt{\beta}$ and by choosing $\beta_0$ as the value of $\beta$ minimizing it.} so it is desirable to increase $\beta$ until \eqref{eq:gen_norm_cons} is fulfilled with equality. We could thus formulate our problem as
\begin{equation}\label{eq:bet_prob}
    \max_{\widetilde{\bs{U}}} \beta , \;\;    \mrm{s.t. } \Vert \bs{\Theta}_r \Vert_2^2 = 1.
\end{equation}
{ However, for some parameter combinations, specifically when the direct channel $\bs{H}_0$ has strong presence, it may not be even possible to fulfill the constraint in \eqref{eq:bet_prob}. Hence, a prerequisite for even attempting to solve \eqref{eq:bet_prob} is to find an initial point where we can have channel orthogonalization while fulfilling the respective RS passive constraint. To this end, our proposed solution considers the initial problem of minimizing $\Vert \bs{\Theta}_r \Vert_2^2$ over $\beta$ and $\widetilde{\bs{U}}$, and use the result as a starting point for attempting to solve  \eqref{eq:bet_prob}. If such initial point does not exist, we can conclude that channel orthogonalization is not possible with a passive RS, and we would need amplification to compensate for the direct channel.} 

In the rest of this section, we will use the general formulation of $\bs{\Theta}_r$ from \eqref{eq:theta_gen} to be able to provide general results applicable to the three RS models under study. However, the fact that $\bs{Z}_S$ and $\bs{Z}_{U/L}$ have several zero columns, as well as the properties of the commutation matrix, allow for some improvement in computation efficiency for ARIS and BD-RIS which may be leveraged for the numerical results. In the case of FRIS, we could also consider the alternative expression for the reflection matrix given in \eqref{eq:FRIS_conf_alt} to improve this efficiency, but this requires the knowledge of $\bs{H}_1$ and $\bs{H}_2$ (up to a shared scalar). We will see that, unlike in \cite{globecom22} where the FRIS is allowed to transmit pilots, a passive FRIS only allows us to estimate $\bs{\mathcal{H}}$. However, Kronecker product decomposition methods \cite{kron_dec} could be potentially employed on $\bs{\mathcal{H}}$ to obtain estimates of $\bs{H}_1$ and $\bs{H}_2$, up to shared scaling.

\subsection{RS power minimization for initialization}
The initial concern is to find if a combination of $\beta$ and $\widetilde{\bs{U}}$ such that \eqref{eq:gen_norm_cons} can be fulfilled. We may tackle this by minimizing $\Vert\bs{\Theta}_r\Vert^2_2$ with respect to $\widetilde{\bs{U}}$ and $\beta$ and checking if the minimum value fulfills said constraint. The spectral norm $\Vert\bs{\Theta}_r\Vert^2_2$, given by the largest singular value of $\bs{\Theta}_r$, is generally difficult to minimize. On the other hand, the Frobenius norm $\Vert \bs{\Theta}_r \Vert_\mrm{F}=\mrm{tr}(\bs{\Theta}_r^\mrm{H}\bs{\Theta}_r)$ corresponds to an equivalent matrix norm which upper-bounds the spectral norm, so by minimizing the Frobenius norm we can also reduce the spectral norm to a great extent. Let us thus consider the problem of minimizing the Frobenius norm $\Vert \bs{\Theta}\Vert^2_\mrm{F}=\mrm{tr}(\bs{\Theta}^\mrm{H}\bs{\Theta})$, which coincides with the Euclidean norm for vectorized matrices.

Considering \eqref{eq:theta_gen}, the squared Frobenius norm of $\bs{\Theta}_r$ for $r\in \{\mrm{FRIS}, \mrm{BD-RIS}, \mrm{ARIS}\}$ may be expressed as
\begin{equation}\label{eq:gen_Fnorm}
    P_r^\mrm{Fro}(\sqrt{\beta},\widetilde{\bs{U}})\triangleq
    \Vert \bs{\Theta}_r \Vert^2_\mrm{F} = \beta g_r(\widetilde{\bs{U}})-2\sqrt{\beta}f_r(\widetilde{\bs{U}})+\kappa_r,
\end{equation}
where
\begin{subequations}\label{eq:gfc}
    \begin{equation}
g_r(\widetilde{\bs{U}})=\vec(\widetilde{\bs{U}})^\mrm{H}\widetilde{\bs{\mathcal{H}}}^\mrm{H}_r \widetilde{\bs{\mathcal{H}}}_r \vec(\widetilde{\bs{U}})
    \end{equation}
    \begin{equation}
        f_r(\widetilde{\bs{U}})=\mathfrak{Re}\big\{\vec(\widetilde{\bs{U}})^\mrm{H}\widetilde{\bs{\mathcal{H}}}^\mrm{H}_r \widetilde{\bs{\mathcal{H}}}_r \vec(\bs{H}_0)\big\}
    \end{equation}
    \begin{equation}
        \kappa_r = \vec(\bs{H}_0)^\mrm{H}\widetilde{\bs{\mathcal{H}}}_r^\mrm{H} \widetilde{\bs{\mathcal{H}}}_r \vec(\bs{H}_0),
    \end{equation}
\end{subequations}
with $\widetilde{\bs{\mathcal{H}}}_r$ given in \eqref{eq:HH_r} for $r\in \{\mrm{FRIS, BD-RIS, ARIS}\}$. Note that $g_r(\widetilde{\bs{U}})$ and $\kappa_r$ are always positive, while $f_r(\widetilde{\bs{U}})$ can be made positive or negative through $\widetilde{\bs{U}}$. The problem we want to solve may be formulated as
\begin{equation}\label{eq:norm_min}
\begin{aligned}
    \arg \min_{\sqrt{\beta}, \widetilde{\bs{U}}}&\; P_r^\mrm{Fro}(\sqrt{\beta},\widetilde{\bs{U}})\\
    \mrm{s.t.}&\; \widetilde{\bs{U}}^\mrm{H}\widetilde{\bs{U}} = \mathbf{I}_K.
\end{aligned}
\end{equation}
Since this problem is quadratic in $\sqrt{\beta}$, the optimal value for given $\widetilde{\bs{U}}$ corresponds to the stationary point
\begin{equation}\label{eq:bet_opt}
    \sqrt{\beta_\mrm{opt}(\widetilde{\bs{U}})} = \frac{f_r(\widetilde{\bs{U}})}{g_r(\widetilde{\bs{U}})},
\end{equation}
where we may assume that $\sqrt{\beta_\mrm{opt}(\widetilde{\bs{U}})}$ is positive since, from \eqref{eq:gfc}, the sign of $f_r(\widetilde{\bs{U}})$ can be absorbed in $\widetilde{\bs{U}}$ without loss of generality (WLOG).\footnote{If $\widetilde{\bs{U}}$ is a stationary points with unitary constraint, then $-\widetilde{\bs{U}}$ is also a stationary point to the Riemannian gradient structure \cite{traian}.} After substituting $\sqrt{\beta_\mrm{opt}(\widetilde{\bs{U}})}$ in \eqref{eq:gen_Fnorm} we get the equivalent problem
\begin{equation}\label{eq:norm_min_op_bet}
\begin{aligned}
    \arg \max_{\widetilde{\bs{U}}}&\; \frac{f_r^2(\widetilde{\bs{U}})}{g_r(\widetilde{\bs{U}})}\\
    \mrm{s.t.}&\; \widetilde{\bs{U}}^\mrm{H}\widetilde{\bs{U}} = \mathbf{I}_K.
\end{aligned}
\end{equation}
The previous maximization problem is generally non-convex. However, we may reach a local minimum by considering gradient ascent algorithms in the unitary group, as those proposed in \cite{traian}. To this end, we first need to characterize the gradient of the objective function in \eqref{eq:norm_min_op_bet}, given by
\begin{equation}
\bs{\Gamma}(\widetilde{\bs{U}})\triangleq\left(\frac{f_r^2(\widetilde{\bs{U}})}{g_r(\widetilde{\bs{U}})}\right)'\!\!=\frac{2f_r(\widetilde{\bs{U}})g_r(\widetilde{\bs{U}})f_r'(\widetilde{\bs{U}})-f_r^2(\widetilde{\bs{U}})g_r'(\widetilde{\bs{U}})}{g_r^2(\widetilde{\bs{U}})},
\end{equation}
where we have
\begin{subequations}\label{eq:der_f_g}
 \begin{equation}
    f_r'(\widetilde{\bs{U}})\triangleq \frac{\partial f_r(\widetilde{\bs{U}})}{\partial \widetilde{\bs{U}}^*} =\vec^{-1}\big( \widetilde{\bs{\mathcal{H}}}^\mrm{H}_r \widetilde{\bs{\mathcal{H}}}_r \vec(\bs{H}_0)\big)
\end{equation}
\begin{equation}
    g_r'(\widetilde{\bs{U}})\triangleq \frac{\partial g_r(\widetilde{\bs{U}})}{\partial \widetilde{\bs{U}}^*} =\vec^{-1}\big( \widetilde{\bs{\mathcal{H}}}^\mrm{H}_r \widetilde{\bs{\mathcal{H}}}_r \vec(\widetilde{\bs{U}})\big).
\end{equation}
\end{subequations}
We may then employ the algorithm from \cite[Table~II]{traian}, which consists of computing the Riemannian gradient,\footnote{ $\widetilde{\bs{U}}$ may be extended to unitary by including $M-K$ orthonormal columns and multiplying by a cropped identity. However, these columns have no impact on the Riemannian gradient since their Euclidean gradient is $\bs{0}$.} and computing the gradient update using the geodesic equation. The algorithm in \cite[Table~II]{traian} further considers the use of Armijo line search for computing the the step size, which assures better convergence (the algorithm will converge almost surely to a local minimum \cite{polak_opt}).

We further propose a closed-form heuristic selection of $\widetilde{\bs{U}}$ to decrease $\Vert\bs{\Theta}_r(\widetilde{\bs{U}})\Vert^2_\mrm{F}$ which may be used as a starting point for \eqref{eq:norm_min_op_bet}. If we take a look at \eqref{eq:theta_gen}, ignoring the inverse vectorization, we note that we have a product of two terms. The first term is a matrix determined by the cascaded channel, and the second term corresponds to the difference between the direct channel and the desired channel. In light of this, we may consider a strategy of choosing the desired channel such that the second term falls close to the lowest eigenmode of the first term. Note that, in the unrestricted case, fixing the second term such that it falls precisely in the lowest eigenmode leads to the optimal selection with fixed norm. Considering the orthogonal desired channel constraint, we may select 
\begin{equation}\label{eq:U_adhoc}
\widetilde{\bs{U}}=\mathcal{P}_{\mathcal{S}(M,K)}\Big(\vec^{-1}\big(\bs{v}_{r,MK}+\vec(\bs{H}_0)\big)\Big),
\end{equation}
where $\bs{v}_{r,MK}$ is the right unitary vector of $\bs{\mathcal{H}}_r$ associated to the lowest singular value, and $\mathcal{P}_{\mathcal{S}(M,K)}(\cdot)$ is the projection under any unitarily invariant norm (e.g., Frobenius and spectral norm) onto the Stiefel manifold $\mathcal{S}(M,K)$ given by \cite{proj}
\begin{equation}\label{eq:proj}
    \mathcal{P}_{\mathcal{S}(M,K)}(\bs{A})=\bs{U}_{\bs{A}}
    \begin{bmatrix}
        \mathbf{I}_K \\
        \bs{0}
    \end{bmatrix}\bs{V}_{\bs{A}}^\mrm{H},
\end{equation}
where $\bs{U}_{\bs{A}}$ and $\bs{V}_{\bs{A}}^\mrm{H}$ are the left and right unitary matrices of the singular value decomposition of the arbitrary $M\times K$ matrix $\bs{A}$. Note that $\mathcal{P}_{\mathcal{S}(M,K)}(\bs{A})$ corresponds to the rotation matrix associated to the polar decomposition of $\bs{A}$ \cite{polar_svd}. If the direct channel $\bs{H}_0$ is dominant over $\bs{v}_{r,MK}$, the selection will favor minimizing the distance between the desired orthogonal channel and the direct channel, whereas, if the direct channel is strongly blocked, the selection will favor minimizing the distance between the desired orthogonal channel and the lowest eigenmode of $\bs{\mathcal{H}}_r$.

\subsection{Channel gain maximization}
As previously mentioned, we are interested in having the channel gain $\beta$ as high as possible since it has direct impact on the rate at which each of the UEs may transmit data. Hence, assuming that we can find $\beta$ and $\widetilde{\bs{U}}$ such that \eqref{eq:gen_norm_cons} is fulfilled (e.g., by using the proposed approximate solutions to \eqref{eq:norm_min}) the next aim is to maximize $\beta$ by trying to solve \eqref{eq:bet_prob}.

Let us assume that we can find $\widetilde{\bs{U}}_0$ such that, for some $\beta_0\geq\beta_\mrm{opt}(\widetilde{\bs{U}}_0)$ from \eqref{eq:bet_opt}, we get a reflection matrix fulfilling \eqref{eq:gen_norm_cons} with strict inequality, i.e., $\Vert \bs{\Theta}_r\Vert^2_2<1$. It is thus desirable to increase $\beta$ until the passive constraint is fulfilled with equality, i.e., $\Vert \bs{\Theta}_r\Vert^2_2=1$. However, due to the characteristics of the spectral norm, it is not obvious how much $\beta$ can be increased. The reason is that the vector $\bs{x}$ in \eqref{eq:gen_norm_cons}, which we may assume to be a unit vector WLOG, is dependent on $\beta$. To solve this problem, we propose an iterative procedure where, starting with $\beta_0$, we alternate between obtaining the unit vector $\bs{x}_i$ solving the maximization in \eqref{eq:gen_norm_cons} for the current $\beta_i$, and obtaining the highest $\beta_{i+1}$ such that $\bs{x}^\mrm{H}_i\bs{\Theta}_r(\beta_{i+1})^\mrm{H}\bs{\Theta}_r(\beta_{i+1})\bs{x}_i=1$, which corresponds to solving a quadratic equation in $\sqrt{\beta_{i+1}}$. The convergence of this method is assured by observing that $\{\Vert \bs{\Theta}_r(\beta_i)\Vert^2_2\}_{i\geq 1}$ is a decreasing sequence lower bounded by $1$, while we know there exists $\beta_0$ such that $\Vert \bs{\Theta}_r(\beta_0)\Vert^2_2<1$. Hence, $\Vert \bs{\Theta}_r(\beta_i)\Vert^2_2$ will converge to 1, and $\beta_i$ will converge to the maximum value allowing for $\Vert \bs{\Theta}_r(\beta_i)\Vert^2_2=1$ with the given $\widetilde{\bs{U}}=\widetilde{\bs{U}}_0$.

{Assuming a feasible point can be found through the RS power minimization used for initialization, the previous method finds $\beta$ such that \eqref{eq:gen_norm_cons} is fulfilled with equality.} We can now try to solve \eqref{eq:bet_prob} by iteratively minimizing  $\Vert \bs{\Theta}_r(\widetilde{\bs{U}})|_{\beta=\beta_i}\Vert^2_2$ over $\widetilde{\bs{U}}$, and use the previous method to increment $\beta$ until $\Vert \bs{\Theta}_r(\beta_i)|_{\widetilde{\bs{U}}=\widetilde{\bs{U}}_i}\Vert^2_2=1$. If we can find the global minimizer $\widetilde{\bs{U}}_i$ of $\Vert \bs{\Theta}_r(\widetilde{\bs{U}})|_{\beta=\beta_i}\Vert^2_2$ at iteration $i$, this method will converge to the optimal solution of \eqref{eq:bet_prob}, since in each iteration $\beta$ is increased and upper bounded by the finite\footnote{Note that in order to keep the constraint \eqref{eq:gen_norm_cons}, $\beta$ must be finite.} optimal value. However, finding a global minimizer of $\Vert \bs{\Theta}_r(\widetilde{\bs{U}})|_{\beta=\beta_i}\Vert^2_2$ is non-trivial. We thus consider approximate solutions by relaxing the spectral norm as before, leading to \eqref{eq:norm_min}, but where $\beta$ is now treated as constant instead of as an optimization variable. This can be solved by using methods for minimization with unitary constraints as the ones considered for solving \eqref{eq:norm_min_op_bet}. Specifically, we consider again the algorithm from \cite[Table~II]{traian}, where the gradient of the objective function is now given by
\begin{equation}
\bs{\Gamma}_{\beta}(\widetilde{\bs{U}})\triangleq \frac{\partial P_r^\mrm{Fro}(\sqrt{\beta},\widetilde{\bs{U}})}{\partial\widetilde{\bs{U}} }=\beta g_r'(\widetilde{\bs{U}})-2\sqrt{\beta}f_r'(\widetilde{\bs{U}}),
\end{equation}
where $g_r'(\widetilde{\bs{U}})$ and $f_r'(\widetilde{\bs{U}})$ are given in \eqref{eq:der_f_g}.

Algorithm~\ref{alg:ch_sel} summarizes the proposed method for orthogonal channel selection. We have used the notation $\bs{v}_{\max}(\cdot)$ to denote an eigenvector associated to the largest eigenvalue of the positive semi-definite input matrix. A simplified orthogonal channel selection can be performed by using directly \eqref{eq:U_adhoc}, leading to a substantial reduction in computation time.

 \begin{algorithm}
 \caption{Orthogonal channel selection algorithm.}
 \label{alg:ch_sel}
 \begin{algorithmic}[1]
 \renewcommand{\algorithmicrequire}{\textbf{Input:}}
 \renewcommand{\algorithmicensure}{\textbf{Output:}}
  \REQUIRE $\bs{H}_0$, $\widetilde{\bs{\mathcal{H}}}_r$
  \ENSURE $\beta$, $\widetilde{\bs{U}}$
  \\ \textit{RS power minimization for initialization}:
  \STATE Select initial $\widetilde{\bs{U}}$ and $\beta$, e.g., by \eqref{eq:U_adhoc} and \eqref{eq:bet_opt}.
 \IF{$\Vert\bs{\Theta}_r\Vert^2_2>1$}
 \STATE Use \cite[Table~II]{traian} to find $\widetilde{\bs{U}}$ maximizing \eqref{eq:norm_min_op_bet}
 \STATE $\beta \leftarrow$ Update with \eqref{eq:bet_opt}
 \IF{$\Vert\bs{\Theta}_r\Vert^2_2>1$}
\STATE Active amplification may be required
\ENDIF
\ENDIF
\\ \textit{Channel gain maximization}:
\WHILE{$\beta$ has not converged}
\STATE Use \cite[Table~II]{traian} to find $\widetilde{\bs{U}}$ minimizing \eqref{eq:gen_norm_cons} (fixed $\beta$)
\WHILE{$\beta$ has not converged}
\STATE $\bs{x}=\bs{v}_{\max}\big(\bs{\Theta}_r^\mrm{H}(\beta)\bs{\Theta}_r(\beta)\big)$
\STATE $\sqrt{\beta}\leftarrow$ Largest root of $\bs{x}^\mrm{H}\bs{\Theta}_r^\mrm{H}(\beta)\bs{\Theta}_r(\beta)\bs{x}-1$
 \ENDWHILE
 \ENDWHILE
  \end{algorithmic}
 \end{algorithm}

\section{Channel estimation and RS configuration}\label{sec:ch}
We now turn our focus into studying the estimation of the channel coefficients required to be able to apply the respective configuration within the different RS models considered. The aim is to characterize the overhead required to obtain the parameters that allow performing channel selection and RS configuration. In \eqref{eq:theta_gen}, we have a general expression for the desired reflection matrix of the RS technologies under study. We thus need to estimate the channel parameters that allow characterizing $\widetilde{\bs{\mathcal{H}}}_r$ and $\bs{H}_0$, which are the only parameters employed in the computation of the channel selection and RS configuration. From \eqref{eq:HH_r}, the required channel parameters are given by $\bs{\mathcal{H}}$, $(\bs{\mathcal{H}}_\mrm{U}+\bs{\mathcal{H}}_\mrm{L})$, and $\bs{\mathcal{H}}_\mrm{S}$ for FRIS, BD-RIS, and ARIS, respectively.\footnote{The remaining matrices in \eqref{eq:HH_r} are fully determined by $M$, $K$, and $N$.} Note that, $(\bs{\mathcal{H}}_\mrm{U}+\bs{\mathcal{H}}_\mrm{L})$ and $\bs{\mathcal{H}}_\mrm{S}$ correspond to reduced matrices coming from $\bs{\mathcal{H}}$ by selecting and/or combining columns, so BD-RIS and ARIS may be seen as a special case of FRIS with restricted channel knowledge.

Due to the passive (energy-efficient) nature of RS technology\cite{RIS_vs_relay,RIS_survey}, we assume that, during the training phase, the estimation/computation tasks are carried out at the BS, while the RS only needs to configure its impedance networks by following a preprogrammed training sequence. Once the final RS configuration is computed at the BS, it is then forwarded to the RS which may then implement it by tuning its impedance networks accordingly. We also assume that the desired channel is conveniently selected at the BS (e.g., through the proposed channel selection algorithm) so that it can use it in the data phase for decoding/precoding purposes without the need for extra channel estimation steps. Next, we describe a potential estimation and configuration scheme consisting of three stages: direct channel estimation, cascaded channel coefficients estimation, and RS configuration.

\subsection{Direct Channel Estimation}
Assuming the direct channel $\bs{H}_0$ is not blocked, all the considered RS models require full knowledge of said channel. Hence, in the initial step the RS would configure its reflection matrix $\bs{\Theta}=\bs{0}$, which is allowed in the FRIS, ARIS, and BD-RIS models. This corresponds to configuring the impedance networks such that all the impinging power is dissipated in the RS resistive components. A practical alternative in more restricted RS models is to reflect the power in a direction away from the BS (i.e., putting $\bs{\Theta}$ in the null-space of $\H_1$ and/or $\H_2$) or to use a two step-method with reflection matrices $\bs{\Theta}_1$ and $\bs{\Theta}_2$ such that $\bs{\Theta}_1+\bs{\Theta}_2=\bs{0}$. Assuming $\bs{\Theta}=\bs{0}$ is effectively achieved, the UEs may send a sequence of $K$ orthogonal pilots to estimate $\H_0$. The received signal over the $K$ time slots is a $K\times K$ matrix given by
\begin{equation}\label{eq:RISAest_Y1}
    \bs{Y}_0 = \bs{H}_0\bs{P}+ \bs{N}_0,
\end{equation}
where $\bs{P}$ is the $K\times K$ known pilot matrix fulfilling $\bs{P}\bs{P}^\mrm{H}=E_s \mathbf{I}_K$, and $\bs{N}_0$ is the noise matrix with IID entries $\{\bs{N}_0\}_{i,j}\sim \mathcal{CN}(0,N_0)$. We can then estimate $\bs{H}_0$ by removing the pilot matrix, i.e., multiplying $1/E_s\bs{P}^\mrm{H}$ from the right, which would not affect the distribution of the estimation noise except for the corresponding scaling.

\subsection{Cascaded Channel Coefficients Estimation}
We proceed with the estimation of the required cascaded channel coefficients, associated to matrix $\widetilde{\bs{\mathcal{H}}}_r$ from \eqref{eq:HH_r}. In general, it is enough to perform the estimation of the cascaded channel coefficients obtained from a sequence of configurations of the reflection matrix $\bs{\Theta}$ forming a basis for the vector spaces defined by its constraints, i.e., \eqref{eq:FRIS_cons}, \eqref{eq:BD-RIS_cons}, and \eqref{eq:ARIS_cons} for the respective RS models. This estimation would thus capture enough parameters to exploit the whole DoFs of the RS reflection. This can be understood by looking at the linear equation \eqref{eq:FRIS_lineq}, which captures the ability of FRIS to configure the channel, and where the particularization to BD-RIS and ARIS corresponds to imposing the extra constraints on $\vec(\bs{\Theta})$. We will exemplify how this process looks like by considering a simple basis for each the three RS models under study, but extension to other bases is trivial.

\subsubsection{FRIS} This corresponds to the most general case since $\bs{\Theta}_\mrm{FRIS}$ is unrestricted (except for the power constraint). However, all the columns of $\bs{\mathcal{H}}$ have to be estimated to be able to exploit the full capabilities of FRIS. We would thus need a sequence of $N^2$ steps, where in step $n$ the FRIS selects $\bs{\Theta}=\bs{E}_{i_n,j_n}$, with $i_n$ and $j_n$ corresponding to the indexes such that $\vec(\bs{\Theta})$ has its non-zero entry at position $n$ (leading to the canonical basis for $\vec(\bs{\Theta}))$.\footnote{In practical scenarios it may be more interesting to select a basis such that entries of $\bs{\Theta}$ selected as 0 are substituted instead by $-1$ so that more power is being reflected, hence avoiding receiver sensitivity issues. This could even be extended to the estimation of $\bs{H}_0$ if use the two step method previously mentioned employing two $\bs{\Theta}$ configurations adding to $\bs{0}$.} At each step, the UEs would send $K$ orthogonal pilots, leading to the following received matrix at step $n$
\begin{equation}\label{eq:est_FRIS}
\bs{Y}_n = \bs{H}_0\bs{P}+  \bs{h}_{1i_n}\bs{h}_{2j_n}^T\bs{P}+ \bs{N}_n,
\end{equation}
where $\bs{h}_{1i_n}$ is the $i_n$th column of $\bs{H}_1$, $\bs{h}_{2j_n}^\mrm{T}$ is the $j_n$th row of $\bs{H}_2$, and $\bs{N}_n$ is the corresponding IID Gaussian noise matrix. In \eqref{eq:est_FRIS}, we can remove the pilot matrix as before and subtract our estimate of $\bs{H}_0$ from the previous stage, leading to an estimate of $\bs{h}_{1i_n}\bs{h}_{2j_n}^T$ with IID Gaussian estimation noise (since we have just applied unitary transformations and combined IID Gaussian matrices). We may note that $\vec(\bs{h}_{1i_n}\bs{h}_{2j_n}^T)$ corresponds to the $n$th column of $\bs{\mathcal{H}}$, so, after repeating the process for the sequence of $N^2$ basis configurations, we would reach an estimate of the whole $\bs{\mathcal{H}}$. {A reduced channel estimation could also be performed by discarding the subset of $\vec(\bs{\Theta})$ falling in the right null-space of $\bs{\mathcal{H}}$. Without any underlying assumptions on $\bs{\mathcal{H}}$, this could be achieved by, e.g., fixing $N^2-MK$ entries of $\vec(\bs{\Theta})$, leading to a reduced number of $MK$ esimation steps. However, these entries should be also discarded when employing FRIS for channel selection, e.g., we could have some FRIS elements not connected to each other. Hence, this would reduce the DoFs of FRIS, which would limit its performance gain over ARIS.}

\subsubsection{BD-RIS} The constraint on $\bs{\Theta}$ of being symmetric now translates into having only $N(N+1)/2$ free variables. Note that $\vec(\bs{\Theta})$ in \eqref{eq:FRIS_lineq} would have $N(N-1)/2$ repeated entries. Alternatively, we could look at the isolated DoFs captured by $\bs{\phi}$ in \eqref{eq:BD-RIS_lineq}. We can proceed as in the FRIS case, but we would now have $N(N+1)/2$ steps, where in step $n$ we may configure the RS with reflection matrix $\bs{\Theta}=\bs{E}_{i_n,j_n}+\bs{E}_{j_n,i_n}$, generating an orthonormal basis for the space of $N\times N$ symmetric matrices. Again, other less sparse bases may be more desirable to avoid receiver sensitivity issues, but this basis gives a convenient example. The received matrix over $K$ orthogonal pilot slots can now be written as
\begin{equation}
\bs{Y}_n = \bs{H}_0\bs{P}+  \bs{h}_{1i_n}\bs{h}_{2j_n}^T\bs{P}+ \bs{h}_{1j_n}\bs{h}_{2i_n}^T\bs{P}+\bs{N}_n.
\end{equation}
Following the same reasoning as for FRIS, we may reach an estimate of $\bs{h}_{1i_n}\bs{h}_{2j_n}^T+ \bs{h}_{1j_n}\bs{h}_{2i_n}^T$ with IID Gaussian estimation noise. The vectorization of said estimated matrix corresponds to the $n$th column of $(\bs{\mathcal{H}}_\mrm{U}+\bs{\mathcal{H}}_\mrm{L})$, so after $N(N+1)/2$ we would have estimated all the channel coefficients required to configure the BD-RIS. {As happened with FRIS, we could reduce the number of estimation steps at the cost of some performance loss by ignoring some DoF, e.g., fixing $N(N+1)-MK$ entries of $\bs{\phi}$ to 0 in \eqref{eq:BD-RIS_lineq}. This would again reduce the number of estimation steps to $MK$, but we would not be able to completely exploit the benefits of BD-RIS.}

\subsubsection{ARIS} The constraint now is that $\bs{\Theta}$ has to be diagonal. This means that $\vec(\bs{\Theta})$ in \eqref{eq:FRIS_lineq} has only $N$ non zero elements multiplying the respective columns of $\bs{\mathcal{H}}$. We may then follow the same steps as in FRIS, but configuring $\bs{\Theta}=\bs{E}_{n,n}$ at step $n$, which generates an orthonormal basis for the space of $N\times N$ diagonal matrices. Following the same reasoning as in FRIS, we may get an estimate of $\bs{h}_{1n}\bs{h}_{2n}^T$, which corresponds to the $n$th column of $\bs{\mathcal{H}}_\mrm{S}$ in \eqref{eq:ARIS_conf}. Hence, after $N$ steps we have estimated all the necessary channel coefficients to configure the ARIS, where the estimation noise is again IID Gaussian.

{If we select the minimum $N$ in the different RS models (i.e., fulfilling with equality \eqref{eq:cond_FRIS}, \eqref{eq:cond_BD-RIS}, and \eqref{eq:cond_ARIS}, respectively) we obtain the minimum number of pilot slots of length $K$ required to estimate the cascaded channel channel parameters for the presented schemes, which gives
\begin{subequations}\label{eq:min_pilots}
    \begin{equation}\label{eq:min_pilots_ARIS}
        L_\mrm{ARIS,min} = MK
    \end{equation}
    \begin{equation}\label{eq:min_pilots_FRIS}
    \begin{aligned}
        L_\mrm{FRIS,ach} &= \max(M,K)^2\\
        &= M^2
    \end{aligned}
    \end{equation}
    \begin{equation}\label{eq:min_pilots_BD-RIS}
    \begin{aligned}
        L_\mrm{BD-RIS,ach} &= \frac{(M+K-1)(M+K-2)}{2}\\
        &= MK+\frac{M(M-3)}{2}+\frac{K(K-3)}{2}+2.
    \end{aligned}
    \end{equation}
\end{subequations}
If no specific structure is assumed on the channel matrices $\bs{H}_1$, and $\bs{H}_2$, we can claim that \eqref{eq:min_pilots_ARIS} formally provides the minimum number of ($K$-sized) pilot slots required to perform channel orthogonalization (or arbitrary channel selection) with ARIS under minimum $N$, i,e, fulfilling \eqref{eq:cond_ARIS} with equality. Thus, the estimation method proposed for ARIS is most efficient in achieving its complete DoFs with minimimum number of pilot resources (under minimum $N$).

In the case of FRIS and BD-RIS, the respective numbers of estimation steps, \eqref{eq:min_pilots_FRIS} and \eqref{eq:min_pilots_BD-RIS}, correspond to achievable bounds, because the required free variables exceed the dimension of the space defined by \eqref{eq:FRIS_cons} and \eqref{eq:BD-RIS_cons}. Thus, it may be possible to have some reductions by more effective estimation methods. We have also presented reduced estimation methods for FRIS and BD-RIS which allow having the same number of estimation steps as ARIS by fixing some entries of their reflection matrices to 0. In fact, it is also trivial to note from \eqref{eq:FRIS_cons} and \eqref{eq:BD-RIS_cons} that $L_\mrm{ARIS,min}$ gives a lower bound for the minimum pilot slots required for FRIS and BD-RIS to achieve arbitrary (orthogonal) channel selection. Thus, the reduced channel estimation methods previously mentioned achieve minimum number of estimation step for FRIS and BD-RIS. Hence, we may also write 
\begin{subequations}
\begin{equation}
    L_\mrm{FRIS,min} = MK
\end{equation}
\begin{equation}
    L_\mrm{BD-RIS,min} = MK
\end{equation}
\end{subequations}
However, for FRIS and BD-RIS, achieving said bound corresponds to sacrificing DoFs which can be otherwise employed to achieve a better overall system performance, e.g., in terms of channel gain. A deep study this trade-off, as well as an investigation of more advanced methods to reduce the estimation steps of FRIS and BD-RIS without sacrificing DoFs, constitute interesting directions for future work.}

\subsection{RS Configuration}
After having estimated the necessary channel parameters to configure the respective RS models, the BS can select a suitable desired channel, e.g., an orthogonal channel with high gain that does not require RS amplification. The BS may hereby employ Algorithm~\ref{alg:ch_sel} for channel selection, which requires only knowledge of the estimated channel parameters (from which we can obtain estimates of $\widetilde{\bs{\mathcal{H}}}_r$ and $\bs{H}_0$). Then, the BS can compute the reflection matrix to be applied at the RS for achieving said channel. The reflection matrix for FRIS, BD-RIS, and ARIS may be computed by using equations \eqref{eq:FRIS_conf}, \eqref{eq:BD-RIS_conf}, and \eqref{eq:ARIS_conf}, respectively. We may further translate the respective reflection matrices into impedance network parameters by considering \eqref{eq:ref_imp}.

\section{Numerical results}\label{section:num_res}
\subsection{Performance under IID Rayleigh fading}
We start by considering a rich multipath propagation environment where $\H_0$, $\H_1$, and $\H_2$ may be modeled as IID Gaussian matrices \cite{mimo}. From the theoretical results in Section~\ref{sec:min}, the ratio between the power of the direct channel and the power of the cascaded channel seems to be a limiting factor for achieving channel orthogonalization. We will thus consider different values for this ratio by fixing the average power of the cascaded channel entries to 1, i.e., $\mathbb{E}\big\{\vert\{\H_1\}_{i,j}\vert^2\big\}=1$ and $\mathbb{E}\big\{\vert\{\H_2\}_{i,j}\vert^2\big\}=1$, and selecting $\eta$ as the power of the direct channel elements, i.e., $\big\{ \vert \{\H_0\}_{i,j}\vert^2\big\}=\eta$, which offers some interplay over said ratio. {The total channel power for blind RS reflection $\bs{\Theta}=\mathbf{I}_N$ (or random with $\mathbb{E}\{\bs{\Theta}\bs{\Theta}^\mrm{H}\}=\mathbf{I}_N$) is thus normalized to $\mathbb{E}\big\{\Vert\{\H\}\Vert^2_{\mrm{F}}\big\}=M K(N+\eta)$.  We should note that, if we select instead $\mathbb{E}\big\{\vert\{\H_1\}_{i,j}\vert^2\big\}=c$ and $\mathbb{E}\big\{\vert\{\H_2\}_{i,j}\vert^2\big\}=1/c$, the presented results would be unaffected since these two channels always appear multiplied. However, in more practical settings it may be required to have the RS closer to the BS to achieve a given $\eta$ ratio \cite{ris_myths}, leading to cascaded channels $\H_1$ and $\H_2$ having different power.}

\begin{figure*}[h]
\begin{subfigure}{\columnwidth}
\centering
    \includegraphics[scale=0.5]{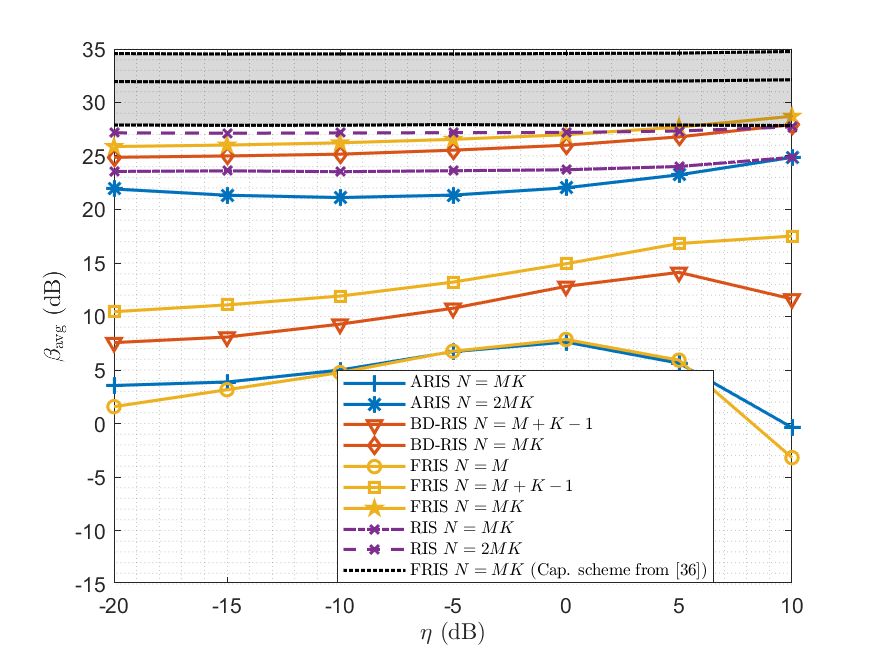}
    \caption{Simplified channel selection from \eqref{eq:U_adhoc}.}
    \label{fig:bet_IID_simp_ch_sel}
  \end{subfigure}
  \begin{subfigure}{\columnwidth}
  \centering
    \includegraphics[scale=0.5]{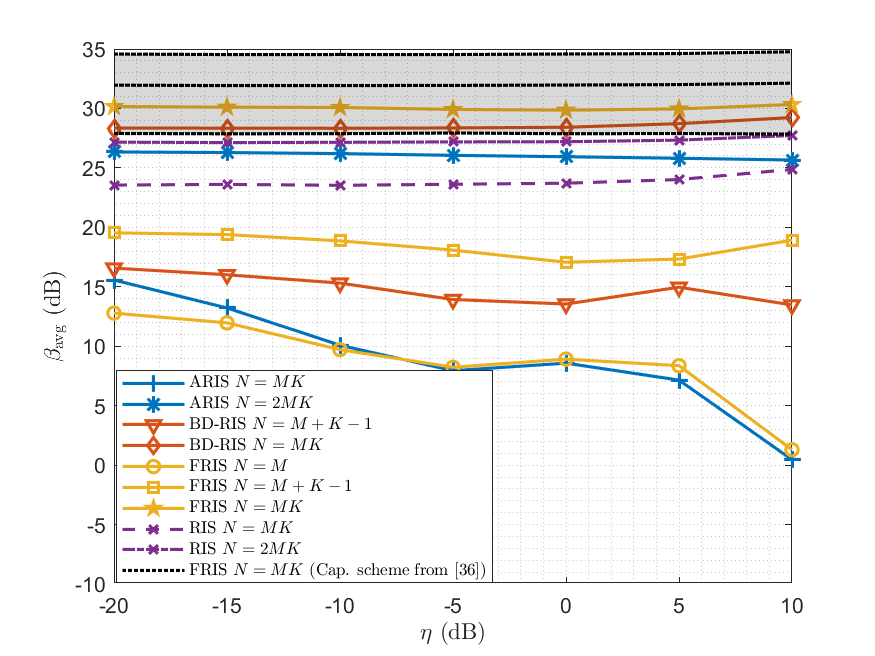}
    \caption{Channel selection through Algorithm~\ref{alg:ch_sel}.}
    \label{fig:bet_IID_ch_sel}
\end{subfigure}
  \caption{Average channel gain versus direct channel power for IID Rayleigh fading scenario with $M=8$ and $K=4$.}
  \label{fig:bet_IID}
  \end{figure*}

In Fig.~\ref{fig:bet_IID}, we evaluate the average channel gain obtained with the different RS technologies after employing Algorithm~\ref{alg:ch_sel} for orthogonal channel selection (Fig.~\ref{fig:bet_IID_ch_sel}), as well as after employing the simplified approach (Fig.~\ref{fig:bet_IID_simp_ch_sel}), which selects \eqref{eq:U_adhoc} directly (i.e., without iterative geodesic gradient descent methods). We have averaged over $10^3$ channel realizations with fixed $M=8$, $K=4$, and varying number of RS elements. Moreover, we have limited the number of iterations of the while loops from \cite[Table~2]{traian} to avoid unreasonable computation times when the convergence is too slow at the cost of limiting the optimality of the results. We have further considered the worst case assumption that the channel gain is 0 whenever the obtained RS configuration does not fulfill the passive constraint. We have also included for comparison a channel orthogonalization approach with RIS, which has been numerically optimized for minimum channel condition number. In this case, $\beta$ represents the average channel gain per UE, but the spread of it is practically unnoticeable, hinting that the orthogonality condition is still practically satisfied. We have also included a baseline approach consisting of the capacity achieving FRIS scheme presented in \cite{bartoli}, which assumes full-blockage of the direct channel for its derivations. In this case, the channel gain is computed as the average of the channel eigenvalues, but we have also illustrated the spread between the average highest and lowest channel eigenvalues as a shaded area around the average. Note that this scheme serves as a performance upper bound when the direct channel is not dominant, but to be able to practically experience these gains at the UEs they would have to jointly collaborate with the BS to diagonalize the channel, which is impractical. The proposed methods, on the other hand, only require of a simple MRC equalizer at the BS to ensure equal channel gain $\beta$ for all UEs, with a respective user rate of $\log (1+\beta \mrm{SNR})$.

\begin{figure*}
  \begin{subfigure}{\columnwidth}
  \centering
    \includegraphics[scale=0.5]{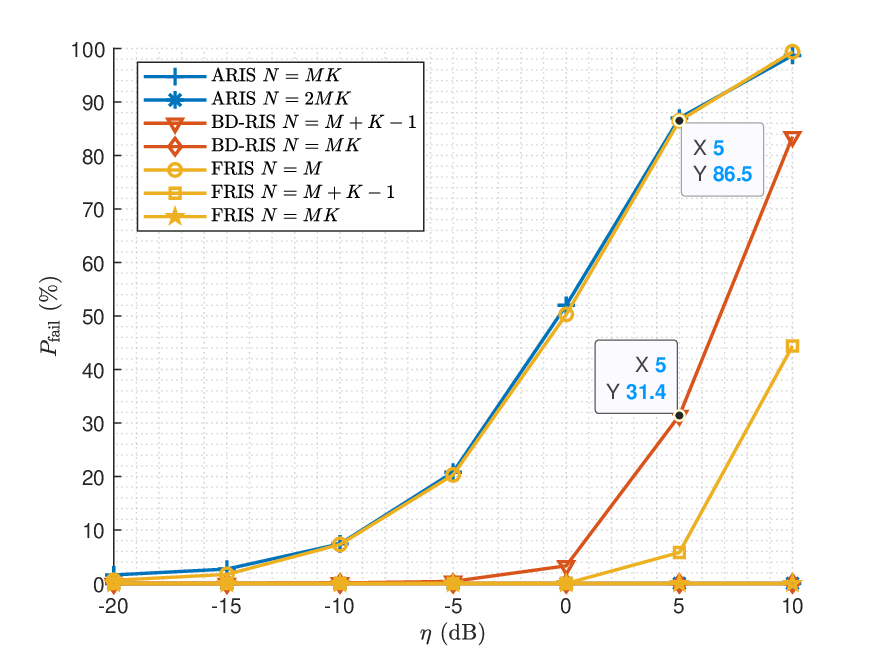}
    \caption{Simplified channel selection from \eqref{eq:U_adhoc}.}
    \label{fig:P_fail_iid_simp}
  \end{subfigure}
  \begin{subfigure}{\columnwidth}
  \centering
    \includegraphics[scale=0.5]{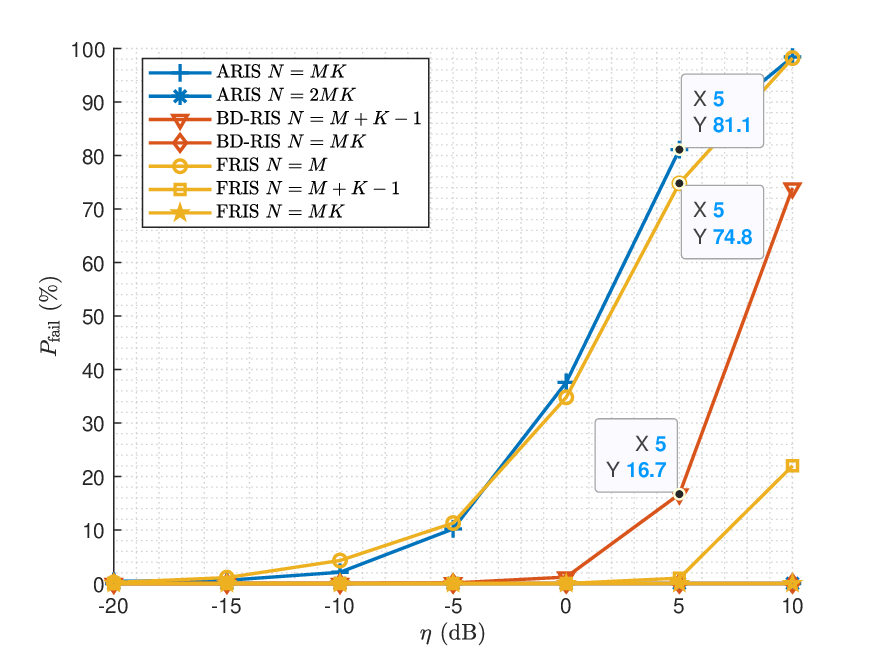}
    \caption{Channel selection through Algorithm~\ref{alg:ch_sel}.}
    \label{fig:P_fail_iid_ch_sel}
  \end{subfigure}
  \caption{Orthogonalization failure probability versus direct channel power for IID Rayleigh fading scenario with $M=8$ and $K=4$.}
    \label{fig:P_fail_iid}
    \vspace{-1.5em}
\end{figure*}

{The average channel gain achieved by the simplified channel selection, shown in
Fig.~\ref{fig:bet_IID_simp_ch_sel}, is substatially increased when using Algorithm~\ref{alg:ch_sel}, shown in Fig.~\ref{fig:bet_IID_ch_sel}, outlining the effectivity of Algorithm~1 in optimizing the channel gain. This increase is more noticeable when the considered RS technologies employ minimum number of elements, as derived in Section~\ref{sec:orth}.} {For equal number of RS elements, both plots confirm the intuition that FRIS outperforms BD-RIS, which outperforms ARIS, outlining also the effectiveness of the channel selection approaches for the three RS technologies. However, with minimum number of RS elements, FRIS performs close to ARIS, while BD-RIS offers a significant improvement. This may be associated to the fact that the DoFs required for BD-RIS to enforce arbitrary channels exceed the required dimensions of the equation space, as remarked in the proof of Theorem~\ref{th:BD-RIS}. When comparing to the baseline approaches, we note that the considered RS technologies require some excess of RS elements over the derived minimum ones for the channel selection to be able to be competitive. This could also be associated to the fact that the number of RS elements is directly related to the beamforming gain obtained from the cascaded channel, so increasing this number would also reduce the dominance of the direct channel, and less power would be required to compensate this channel. Nevertheless, for equal number of elements $N=MK$, BD-RIS and FRIS have marginal loss over the capacity scheme from \cite{bartoli}, while they even outperform the average channel gain obtained by the worst UE. ARIS, on the other hand, can approach the capacity achieving scheme by doubling its number of elements, while it is still slightly outperformed by RIS, which can attain higher channel gain by slightly relaxing the orthogonality constraint. However, we will show that the considered RS technologies are also more robust than RIS towards towards maintaining channel orthogonality in the presence of imperfect-CSI. Moreover, the RIS scheme considered here requires knowledge of the complete cascaded channels, and it relies on impractical numerical optimization methods.

Fig.~\ref{fig:P_fail_iid} shows the failure rate $P_\mrm{fail}$, which is defined as the rate of channel realizations where the channel selection leads to a configuration not fulfilling the passive constraint, i.e., giving $\beta=0$ in our previous simulation. These results are obtained from the same simulations leading to Fig.~\ref{fig:bet_IID}. The RIS model is not hereby compared since it always leads to non-zero orthogonality mismatch by enforcing the full-reflection constraint. As expected, the higher the direct channel power the higher the failure rate since more power is required to compensate this channel, increasing the minimum $\beta$ to do it (recall that $\sqrt{\beta}$ is quadratically related to the RS power). This also justifies the performance gap from Fig.~\ref{fig:bet_IID} when the RS technologies use minimum number of elements. On the other hand, adding extra RS elements eventually leads to $0\%$ failure in the considered range, achieved in our results for ARIS with $N=2MK$, or for BD-RIS and FRIS with $N=MK$. This may also be connected to the relation between the number of RS-elements and the direct-to-cascaded channel ratio, since, for a given direct channel gain, increasing the RS-elements would increase the cascaded channel gain due to the extra beamforming gain.} {On the other hand, when the RS technologies have minimum number of elements, Algorithm~\ref{alg:ch_sel} (Fig.~\ref{fig:P_fail_iid_ch_sel}) can reduce significantly the failure probability attained by the simplified channel selection (Fig.~\ref{fig:P_fail_iid_simp}). This fact outlines the effectivity of the geodesic gradient descent \cite[Table~II]{traian} for solving \eqref{eq:norm_min_op_bet}, which is employed in the RS power minimization used for initialization to find an achievable starting point that attains orthogonalization under the passive constraint.}
  
\subsection{Orthogonalization accuracy with Imperfect CSI}
{We next study the effect of imperfect-CSI on the accurateness of the channel orthogonalization, which is measured in terms of the condition number of the orthogonalized channel.\footnote{In isotropic propagation environments (e.g., rich scattering), the condition number is associated to the drift from having orthogonality between the subchannels associated to different users since the more variation there is between the eigenvalues of the channel the more concentrated the energy is around some of its orthogonal subspaces.} The channel condition number is also directly linked to multiplexing performance, since a low condition number corresponds to having higher channel orthogonality and user fairness, i.e., making it easier to distinguish UEs at the BS and serve them equally, while the converse applies for a high condition number \cite{mimo}. Specifically, a channel condition number of 0 dB (orthogonal channel) achieves the best multiplexing performance by equally dividing the sum-rate among all UEs (ensuring fairness), while allowing the BS to decode them in parallel and without interference (by simple MRC). The considered channel orthogonalization approaches may be used to enforce a channel condition number of $0$ dB under perfect-CSI assumption, but it is interesting to see how much the orthogonalization is degraded when accounting for imperfect-CSI.}

\begin{figure}[h]
    \centering
    \includegraphics[scale=0.5]{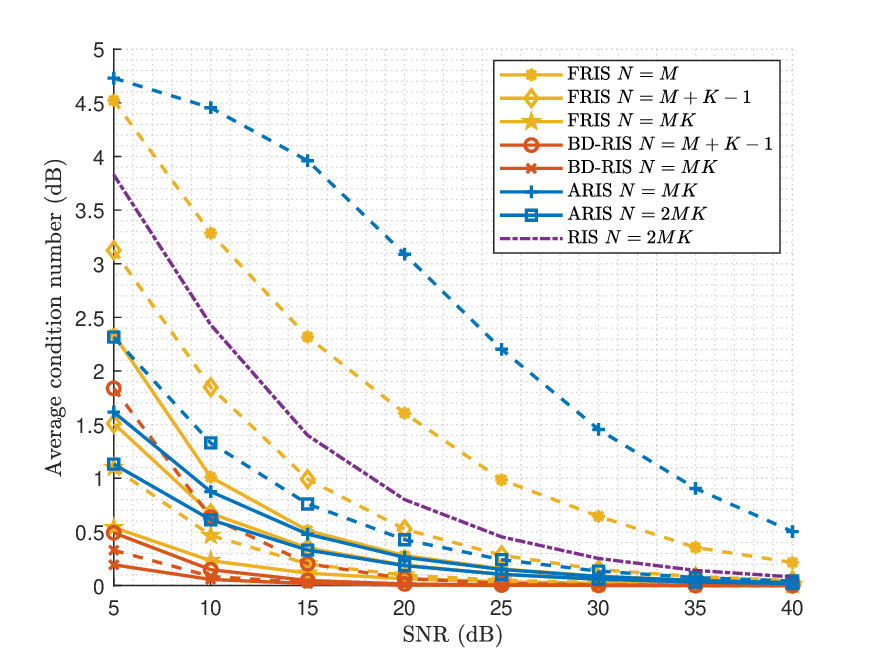}
    \caption{Condition number under imperfect-CSI in MU-MIMO with $M=8$ and $K=4$. Dashed lines indicate random orthogonal channel selection, while solid lines indicate orthogonal channel selection through Algorithm~\ref{alg:ch_sel}.}\label{fig:Imp_CSI}
    \vspace{-1em}
\end{figure}

In Fig.~\ref{fig:Imp_CSI}, we show the condition number of the orthogonalized channel versus channel estimation SNR. We have based the simulations on the channel estimation procedure from Section~\ref{sec:ch}, which considers estimation over a single time-frequency resource, i.e., leading to correspondence between the estimation SNR and the communication SNR. However, we could further increase the estimation SNR by, e.g., combining several time samples or highly-correlated subcarriers. We have also focused on the case where the direct channel is blocked, i.e., $\bs{H}_0=\bs{0}$, since the presence of an imperfect direct channel only leads to an extra addititive IID Gaussian noise term in the final channel, which has simple characterization, while the imperfect-CSI cascaded channel has less predictable impact due to the involvement of complex operations like matrix pseudoinverses. Moreover, $\bs{H}_0=\bs{0}$ assures that the RS models can fulfill the passive constraint in all cases whenever the respective conditions from \eqref{eq:cond_ARIS}, \eqref{eq:cond_BD-RIS}, and \eqref{eq:cond_FRIS} are fulfilled. The simulation averages over $10^3$ IID Gaussian realizations of $\bs{H}_1$ and $\bs{H}_2$ with normalized power (associated to a rich scattering scenario), and we have compared a random orthogonal channel selection (dashed lines) with a channel selection based on Algorithm~\ref{alg:ch_sel} (solid lines). The results showcase how the proposed channel selection is also effective at increasing the robustness of the orthogonalization against CSI imperfections, especially for minimum number of RS elements. Interestingly, BD-RIS outperforms FRIS even for the same number of elements, which can be understood by the fact that the FRIS configuration employs considerably more noisy channel parameters ($N^2$) than BD-RIS ($N(N+1)/2$). All the results show improved orthogonalization robustness of the considered RS technologies over RIS (again numerically optimized for minimum condition number). This is true even with a lower number of elements, or for random orthogonal channel selection with equal number of elements, outlining the importance of sacrificing some reflected power in exchange of accurate channel orthogonalization.

\subsection{Performance under Rician fading scenario}
{We now analyze the spectral efficiency results for a more practical Rician fading indoor scenario depicted in Fig.~\ref{fig:Ric_scen}, where the direct channel between the UEs and the BS may suffer from blockage, and we have $K=3$ UEs randomly located throughout the green area. To account for near-field effects, we consider that only the perpendicular incident power is absorbed by the respective panels \cite{near-field}, while the path-loss is normalized such that the SNR from our results corresponds to the average SNR experienced by the whole BS panel in the absence of the RS and the blockage. The scenario parameters are summarized in Table~\ref{tables:sim_scen}.

\begin{table}[h]
\caption{Rician scenario parameters.}
\centering
\resizebox{0.9\columnwidth}{!}{%
\begin{tabular}{@{}lc@{}}
\toprule
\textbf{Parameter} & \textbf{Value} \\ 
\midrule
Frequency & $3$ GHz \\ 
Rician factor & $5$ dB\\
Room dimensions & $30\lambda \times 30\lambda$
\end{tabular}%
\hspace{1em} \hspace{1em}
\begin{tabular}{@{}lc@{}}
\toprule
\textbf{Parameter} & \textbf{Value} \\ 
\midrule
Antenna spacing &  $\lambda/2$\\
BS panel arrangement & $2\times 2$\\
RS panel arrangement & $2\times 6$
\end{tabular}%
}
\label{tables:sim_scen}
\end{table}

\begin{figure}[h]
    \centering
    \includegraphics[scale=0.5]{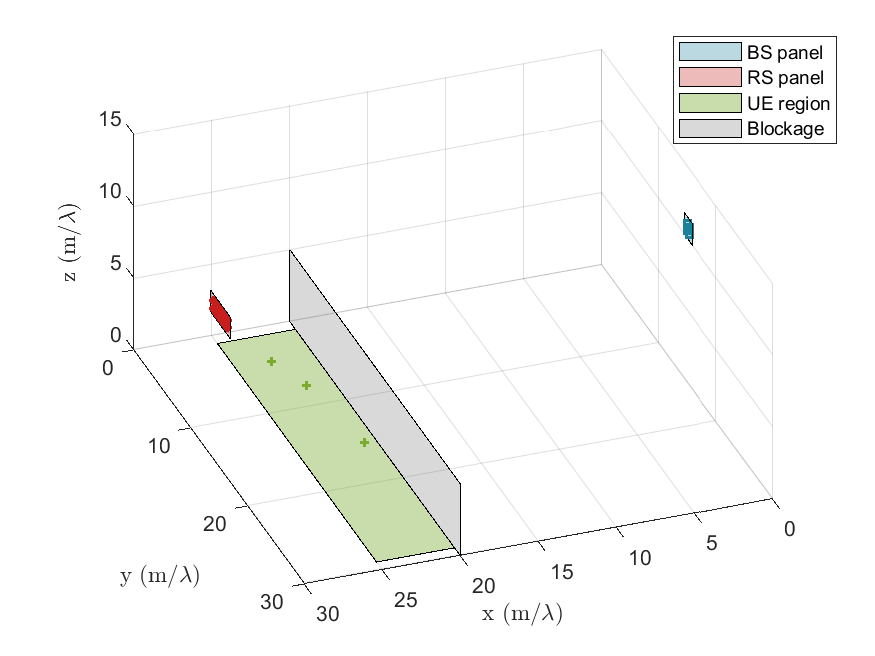}
    \caption{Rician scenario.}\label{fig:Ric_scen}
    \vspace{-1em}
\end{figure}

In Fig.~\ref{fig:ric_rate} we plot the spectral efficiency per UE versus SNR for different levels of direct channel blockage. We have assumed perfect-CSI to ease visualization, but, as hinted from the results in Fig.~\ref{fig:Imp_CSI}, if the estimation-SNR is reasonably high, under imperfect-CSI the curves for ARIS, BD-RIS, and FRIS would only slightly spread around the ones shown in Fig.~\ref{fig:ric_rate}. Note that, due to the enforced orthogonalization with FRIS, BD-RIS, and ARIS, the BS can employ MRC to serve all the UEs with equal spectral efficiency, unlike with the other approaches which show some spread around the average. The results have been averaged over a total of $10^4$ scenarios, consisting of $100$ realizations of the UEs random locations, each considering $100$ realizations of the fading component. As baseline approaches, we have considered the capacity achieving scheme from \cite{bartoli} with MRC processing at the BS, as well as an MRC and a ZF scheme applied at the BS assuming there is no RS. In Fig.~\ref{fig:ric_blk0} we can see that the spectral efficiency for FRIS scales similarly as the ZF scheme even though the BS is restricted to perform only MRC, which simplifies the processing task, incurring lower latency \cite{latency}. BD-RIS also scales closer to ZF, outperforming the other baselines, but there may be some extra loss due to some realizations where it is not able to achieve orthogonalization, which are pessimistically assumed to give null spectral efficiency. The capacity achieving scheme from \cite{bartoli} performs poorly in Fig.~\ref{fig:ric_blk0} due to the low level of blockage of the direct channel, since it is designed assuming full blockage. ARIS is not able to achieve channel orthogonalization at this low level of blockage, even though the RS elements equal its minimum requirement. We can conjecture that Algorithm~\ref{alg:ch_sel} fails to find a channel that allows to compensate for the direct channel with a passive ARIS since it may not have enough DoFs to fully exploit the RS beamforming gain. 

As the level of blockage increases in Fig.~\ref{fig:ric_rate}, the approaches not relying on a RS suffer from an overly weak channel, while ARIS starts to gain importance since it has an easier task at achieving channel orthogonalization. In al cases, BD-RIS and FRIS outperform the capacity achieving scheme from \cite{bartoli} for high enough SNR, while they further ensure better user fairness. This is true even in the presence of complete blockage from Fig.~\ref{fig:ric_blk_inf}, where the scheme from \cite{bartoli} is shown to be capacity achieving. However, the desirable restriction of performing MRC at the BS limits the performance of this method, which requires impractical channel diagonalization and waterfilling to exploit its gains. In the presence of complete blockage from Fig.~\ref{fig:ric_blk_inf}, we can also see that ARIS performance scales similarly as those for BD-RIS and FRIS, which hints that the RS structure can be significantly simplified in the task of channel orthogonalizations with high levels of blockage. Interestingly, BD-RIS surpasses FRIS in this scenario, which may hint that the extra structure in the BD-RIS optimization simplifies slightly the optimization task performed by Algorithm~\ref{alg:ch_sel}. A more effective algorithm for solving \eqref{eq:bet_opt} should be explored in future work to reduce these uncertainties.

\begin{figure*}[h]
\begin{subfigure}{\columnwidth}
\centering
    \includegraphics[scale=0.45]{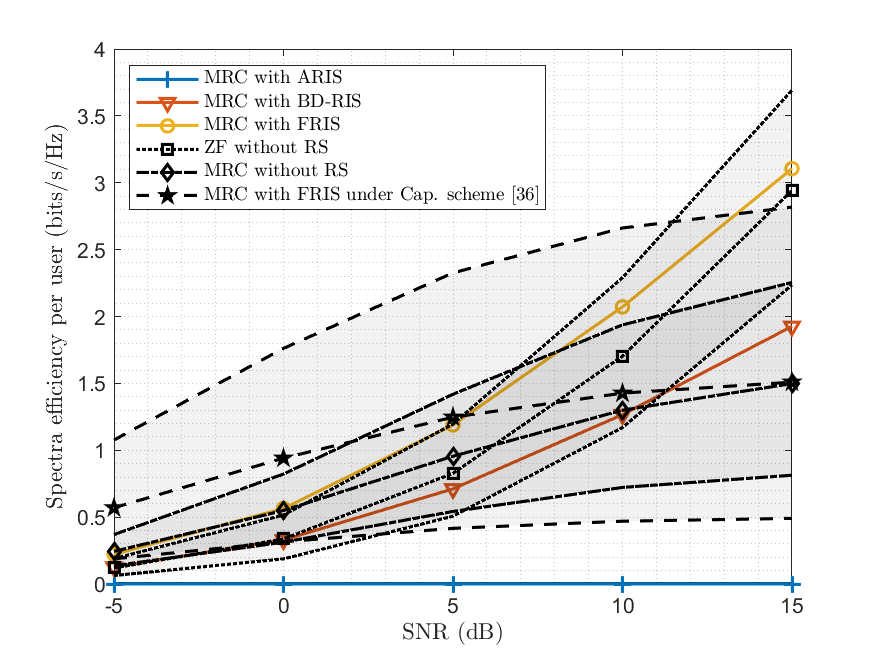}
    \caption{0 dB blockage.}
    \label{fig:ric_blk0}
  \end{subfigure}
  \begin{subfigure}{\columnwidth}
  \centering
    \includegraphics[scale=0.45]{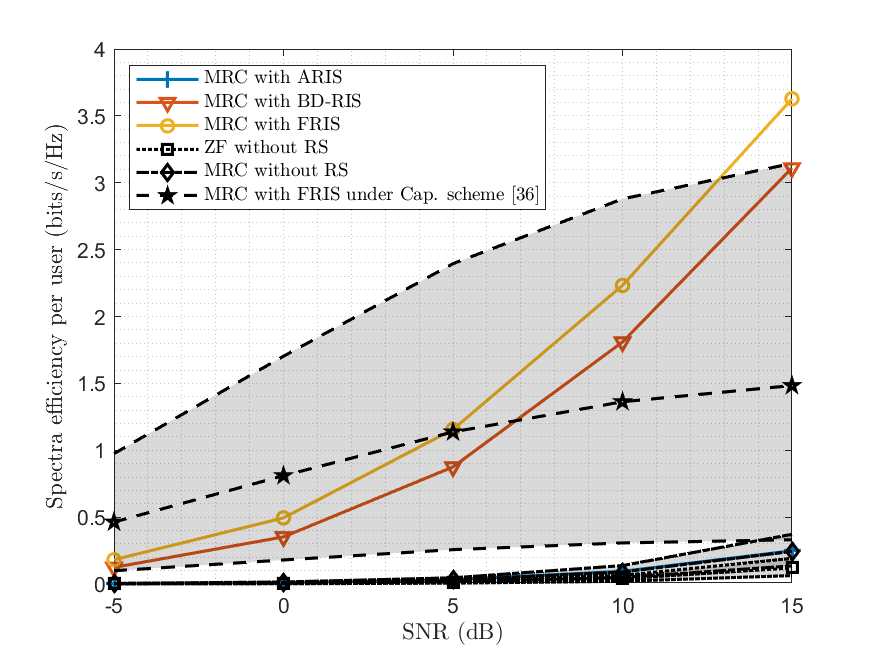}
    \caption{20 dB blockage.}
    \label{fig:ric_blk20}
\end{subfigure}
\begin{subfigure}{\columnwidth}
\centering
    \includegraphics[scale=0.45]{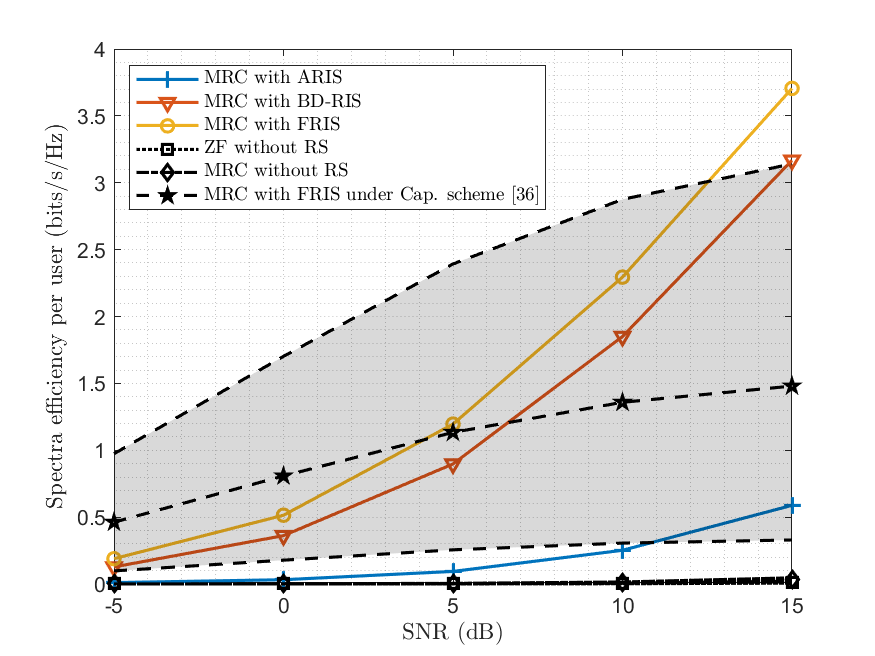}
    \caption{30 dB blockage.}
    \label{fig:ric_blk30}
\end{subfigure}
\begin{subfigure}{\columnwidth}
    \centering
    \includegraphics[scale=0.45]{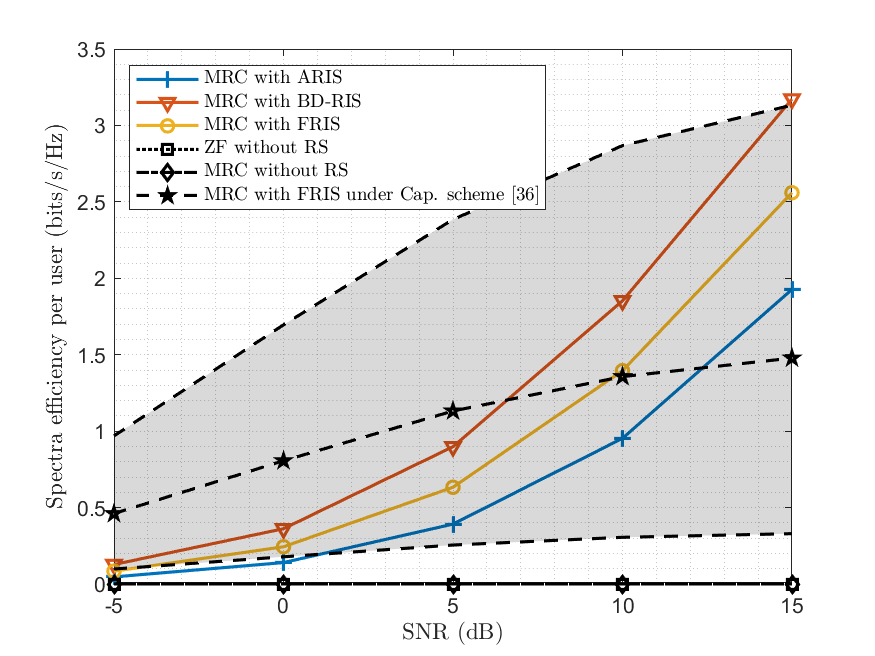}
    \caption{Complete blockage.}
    \label{fig:ric_blk_inf}
\end{subfigure}

  \caption{Spectral efficiency per UE versus SNR for Rician scenario from Fig.~\ref{fig:Ric_scen}. Marked lines indicate the average UE spectral efficiency, while shaded areas between unmarked lines illustrate the spread of UE spectral efficiencies.}
  \label{fig:ric_rate}
      \vspace{-0.5em}
  \end{figure*}
}
\section{Conclusions and Future Work}
We have analyzed the use passive RS technology for channel orthogonalization in MU-MIMO. We have presented three RS models which generalize the widely studied models, and we have discussed their potential implementation, as well as the respective restrictions on the achievable reflection matrices. We have also derived the conditions for achieving arbitrary (orthogonal) channels with the considered RS models. Moreover, we have shown methods to optimize the channel selection procedure, and to practically estimate the channel parameters to be able to apply the corresponding RS configuration. The numerical results have showcased the potential of the presented theory, which allows attaining perfect channel orthogonalization with reduced loss in terms of channel gain. Overall, BD-RIS seems to achieve a good trade-off in terms of performance and complexity, but more research is needed to confirm its practicality.

The presented work takes an important step towards the achievement of OSDM, the spatial counterpart to orthogonal time-frequency modulations as OFDM. However, these results constitute the beginning of a new research direction with many possibilities of extension. For example, future work could consider the use of alternative technologies to RS in the task of channel orthogonalization. Regarding closer extension to the presented work, we could study the use of alternative optimization methods to those considered in Section~\ref{sec:min}. For example, we could consider exploiting the freedom in the pseudoinverse to further improve the results, or we could try to adapt the results to the spectral norm instead of using the relaxation to Frobenius norm. Computational complexity is also a limiting factor, which calls for more advanced methods to avoid matrix inversions and matrix exponentials for the RS configuration and channel selection tasks. Moreover, the imperfect CSI scenario could also be optimized by proposing specific channel selection schemes which take into account the distribution of the estimation error. Another interesting extension could consider studying the interplay between channel gain and level of orthogonality.

The presented results can also be employed to address other research questions. For example, these findings could be employed towards bounding the number of RIS elements required in general scenarios with different goals. On the other hand, the general estimation procedure we studied can also be considered in such scenarios to delimit the required pilot overhead. The theoretical results in Section~\ref{sec:orth} may even find potential application outside the wireless communications context, for example towards analyzing some problems related to control theory.

\appendix[Proof of Theorem~\ref{th:BD-RIS}]
We seek to study the solutions to the matrix equation
\begin{equation}\label{eq:BD-RISeq}
    \bs{H}_0+\bs{H}_1\bs{\Theta}_\mrm{S}\bs{H}_2 = \widetilde{\bs{H}},
\end{equation}
where $\bs{\Theta}_\mrm{S}=\bs{\Theta}_\mrm{S}^\mrm{T}$ with dimension $N\times N$, and $\widetilde{\bs{H}}$ is a non-zero $M\times K$ matrix. We can define WLOG $\bs{\Theta}_\mrm{S}=\bs{\Theta}_\mrm{U/L}+\bs{\Theta}_\mrm{U/L}^\mrm{T}$, where $\bs{\Theta}_\mrm{U/L}$ corresponds to an upper/lower triangular matrix. We can then rewrite \eqref{eq:BD-RISeq} as
\begin{equation}\label{eq:BD-RISeq2}
\bs{H}_1\bs{\Theta}_\mrm{U/L}\bs{H}_2+\bs{H}_1\bs{\Theta}_\mrm{U/L}^\mrm{T}\bs{H}_2+\bs{H}_0 = \widetilde{\bs{H}}.
\end{equation}
After vectorizing, we reach the linear equation
\begin{equation}\label{eq:BD-RIS_lineq}
(\bs{\mathcal{H}}+\bs{\mathcal{H}}\bs{K}^{(N,N)})\bs{Z}_\mrm{U/L}\bs{\phi}=\bs{c},
\end{equation}
where $\bs{\phi}$ is a vector containing the unique unknowns from $\bs{\Theta}_\mrm{S}$ (with the diagonal elements scaled by $1/2$),  $\bs{\mathcal{H}}$ and $\bs{c}$ are defined as in \eqref{eq:FRIS_lineq}, $\bs{K}^{(N,N)}$ is the commutation matrix mapping $\vec(\bs{\Theta}_\mrm{U/L})$ to $\vec(\bs{\Theta}^\mrm{T}_\mrm{U/L})$ \cite{comm_mtx}, and $\bs{Z}_\mrm{U/L}$ is a matrix selecting the columns of $\bs{\mathcal{H}}+\bs{\mathcal{H}}\bs{K}^{(N,N)}$ associated to the $N(N+1)/2$ upper/lower triangular elements after vectorization.\footnote{This matrix may be constructed by, e.g., multiplying $\mathbf{I}_{N^2}$ with the vectorization of a upper/lower triangular matrix with 1s in the upper/lower triangular part.} Let us define the matrix associated to the upper triangular elements as $\bs{\mathcal{H}}_\mrm{U}=\bs{\mathcal{H}}\bs{Z}_\mrm{U}$, while the matrix associated to the lower triangular elements is then given by $\bs{\mathcal{H}}_\mrm{L}=\bs{\mathcal{H}}\bs{K}^{(N,N)}\bs{Z}_\mrm{U}$, since $\bs{Z}_\mrm{L}=\bs{K}^{(N,N)}\bs{Z}_\mrm{U}$ (and vice-versa). The existence of a solution to \eqref{eq:BD-RISeq} is equivalent to the existence of a solution to \eqref{eq:BD-RIS_lineq}, which is given by the rule
\begin{equation}
\rank\big(\bs{\mathcal{H}}_\mrm{U}+\bs{\mathcal{H}}_\mrm{L}\big) =MK.
\end{equation}
Characterizing said rank is not trivial, but, assuming that a solution is available, said solution is given by
\begin{equation}\label{eq:BD-RIS_Tsol}
\bs{\phi}=\big( \bs{\mathcal{H}}_\mrm{U}+\bs{\mathcal{H}}_\mrm{L}\big)^\dagger \bs{c},
\end{equation}
where $(\cdot)^\dagger$ here stands for right pseudoinverse (which may not be unique). The solution to \eqref{eq:BD-RISeq} can be obtained by reorganizing $\bs{\phi}$ into the upper/lower triangular elements of an $N\times N$ matrix, i.e.,
\begin{equation}
    \bs{\Theta}_\mrm{U/L} = \vec^{-1}( \bs{Z}_\mrm{U/L}\bs{\phi}),
\end{equation}
and constructing $\bs{\Theta}_\mrm{S}=\bs{\Theta}_\mrm{U/L}+\bs{\Theta}^\mrm{T}_\mrm{U/L}$, where we can use the commutation matrix to introduce the sum inside the $\vec^{-1}(\cdot)$ operator \cite{comm_mtx}.

We now proceed to prove the conditions under which a solution to \eqref{eq:BD-RISeq} exists. Let us assume \eqref{eq:cond_FRIS} since this is clearly a necessary condition for \eqref{eq:BD-RISeq} due to the stronger constraints on the matrix of unknowns. Given the singular value decomposition $\bs{H}_1=\bs{U}_1\bs{S}_1\bs{V}_1^\mrm{H}$ we can define WLOG $\widetilde{\bs{\Theta}}_\mrm{S}=\bs{V}_1^\mrm{H}\bs{\Theta}_\mrm{S}\bs{V}_1^\mrm{*}$, and multiply from the left both sides of \eqref{eq:BD-RISeq} by $\bs{S}_{1,\mrm{sq}}^{-1}\bs{U}_1^\mrm{H}$, where $\bs{S}_{1,\mrm{sq}}$ is the invertible part of $\bs{S}_1$, to reach
\begin{equation}
    \begin{bmatrix}\mathbf{I}_M & \bs{0}\end{bmatrix}\widetilde{\bs{\Theta}}_\mrm{S}\widetilde{\bs{H}}_2 = \widecheck{\bs{H}},
\end{equation}
where $\widetilde{\bs{H}}_2$ remains a randomly chosen matrix, and we have defined $\widecheck{\bs{H}}=\bs{S}_{1,\mrm{sq}}^{-1}\bs{U}_1^\mrm{H}(\widetilde{\bs{H}}-\bs{H}_0)$. Let us denote the top $M$ rows of $\widetilde{\bs{\Theta}}_\mrm{S}$ by
\begin{equation}\label{eq:johan_trick}
\{\widetilde{\bs{\Theta}}_\mrm{S}\}_{1:M,:} = \begin{bmatrix} \widetilde{\bs{\Theta}}_{11,\mrm{S}} & \widetilde{\bs{\Theta}}_{12}
\end{bmatrix},
\end{equation}
where $\widetilde{\bs{\Theta}}_{11,\mrm{S}}$ must be chosen as a symmetric $M\times M$ matrix, while $\widetilde{\bs{\Theta}}_{12}$ can be chosen as an unrestricted $M\times (N-M)$ matrix. We may rewrite \eqref{eq:johan_trick} as
\begin{equation}
\widetilde{\bs{\Theta}}_{11,\mrm{S}}\widetilde{\bs{H}}_{2,\mrm{T}}+\widetilde{\bs{\Theta}}_{12}\widetilde{\bs{H}}_{2,\mrm{B}}=\widecheck{\bs{H}},
\end{equation}
where $\widetilde{\bs{H}}_{2,\mrm{T}}$ and $\widetilde{\bs{H}}_{2,\mrm{B}}$ correspond to the top $M\times K$ block and the bottom $(N-M)\times K$ block of $\widetilde{\bs{H}}_2$, respectively. We may then apply on $\widetilde{\bs{H}}_{2,\mrm{T}}$ the equivalent trick we used with $\bs{H}_1$ in \eqref{eq:johan_trick}, which here results in
\begin{equation}\label{eq:final_BD-RIS}
\widecheck{\bs{\Theta}}_{11,\mrm{S}}\begin{bmatrix}\mathbf{I}_K & \bs{0}\end{bmatrix}^\mrm{T}=\bar{\bs{H}}-\widecheck{\bs{\Theta}}_{12}\widecheck{\bs{H}}_{2,\mrm{B}}.
\end{equation}
From \eqref{eq:final_BD-RIS} it becomes evident that the solvability is determined by the conditions under which the top $K$ tows of the RHS can be made symmetric. This can be seen from the fact that the left hand side (LHS) of \eqref{eq:final_BD-RIS} corresponds to a $M\times K$ matrix where the bottom $(M-K)\times K$ block can be freely selected, but the top $K\times K$ block has symmetric constraint. Thus, solving \eqref{eq:BD-RISeq} is equivalent to finding an arbitrary $K\times (N-M)$ matrix $\widecheck{\bs{\Theta}}_{12,\mrm{T}}$ such that
\begin{equation}\label{eq:sym_cons}
    \widecheck{\bs{\Theta}}_{12,\mrm{T}}\widecheck{\bs{H}}_{2,\mrm{B}}-\widecheck{\bs{H}}_{2,\mrm{B}}^\mrm{T}\widecheck{\bs{\Theta}}_{12,\mrm{T}}^\mrm{T} = \bar{\bs{H}}_\mrm{T}-\bar{\bs{H}}^\mrm{T}_\mrm{T}.
\end{equation}
If $N-M\geq K$, \eqref{eq:sym_cons} is trivially solvable since this allows to select $\widecheck{\bs{\Theta}}_{12,\mrm{T}}$ such that $\widecheck{\bs{\Theta}}_{12,\mrm{T}}\widecheck{\bs{H}}_{2,\mrm{B}}=\bar{\bs{H}}^\mrm{T}_\mrm{T}$. Let us now assume $N-M< K$. We may now proceed as before to absorb the invertible parts of $\widecheck{\bs{H}}_{2,\mrm{B}}$ in the other matrices until we reach
\begin{equation}\label{eq:fin_BD-RIS_Sym}
    \bar{\bs{\Theta}}_{12}\begin{bmatrix}\mathbf{I}_{N-M} & \bs{0}\end{bmatrix}-\begin{bmatrix}\mathbf{I}_{N-M} \\ \bs{0}\end{bmatrix} \bar{\bs{\Theta}}_{12}^\mrm{T}=\bar{\bar{\bs{H}}}-\bar{\bar{\bs{H}}}^\mrm{T}.
\end{equation}
The LHS then corresponds to a matrix which has a bottom-right block of zeros of dimension $(M+K-N)\times (M+K-N)$, while the other block is anti-symmetric with 0s on the diagonal. Moreover, the RHS is an anti-symmetric matrix with zeros in the diagonal, while the off-diagonals are in general non-zero, leading to the sufficient (``if'') condition \eqref{eq:cond_BD-RIS} associated to having said block of zeros in the LHS of dimension lower or equal to $1$. If we recall the transformations that we have applied to the top $K\times K$ block of $(\widetilde{\bs{H}}-\bs{H}_0)$ until reaching $\bar{\bar{\bs{H}}}$, we have essentially multiplied \textit{randomly chosen} matrices from the right and from the left. Thus, as long as $(\widetilde{\bs{H}}-\bs{H}_0)$ has at least one non-zero entry, the non-diagonal entries of the RHS will be non-zero with probability 1 since they are given by linear combinations of elements of randomly chosen matrices. Hence, \eqref{eq:cond_BD-RIS} is a necessary (``only if'') condition for randomly chosen $\bs{H}_1$ and $\bs{H}_2$, which completes the proof.

%\newpage
\renewcommand{\baselinestretch}{.95}
\bibliographystyle{IEEEtran}
\balance
\bibliography{IEEEabrv,bibliography}

%okay that works
% \fi

\end{document}